\documentclass{article}
\usepackage{graphicx} 
\usepackage{authblk}
\usepackage{adjustbox}
\usepackage{multicol}
\usepackage{multirow}
\usepackage{tabularx}
\usepackage{booktabs}
\usepackage{amsfonts, amsmath, amssymb}
\usepackage[numbers]{natbib}
\usepackage[mathlines]{lineno}
\usepackage{hyperref}
\hypersetup{
    colorlinks=true,
    citecolor=blue,
    filecolor=black,
    linkcolor=blue,
    urlcolor=blue
}
\usepackage{tablefootnote}
\usepackage{makecell}

\usepackage[T1]{fontenc}
\usepackage{lmodern} 

\date{}

\graphicspath{{./Figs/}}

\title{A preference for dynamical phantom dark energy using one-parameter model with Planck, DESI DR1 BAO and SN data}
\author[1]{Ramy Fikri}
\author[1]{Esraa Elkhateeb}
\author[1]{E. I. Lashin}
\author[2]{Waleed El~Hanafy}
\affil[1]{Department of Physics, Faculty of Science, Ain Shams University, Cairo 11566, Egypt}
\affil[2]{Centre for Theoretical Physics, The British University in Egypt, P.O. Box 43, El Sherouk City, Cairo 11837, Egypt}

\begin{document}

\maketitle
\begin{abstract}
Baryon Acoustic Oscillation (BAO) provides a powerful tool to measure cosmic expansion and consequently the nature of the Dark Energy (DE). Recent precise BAO measurements by Dark Energy Spectroscopic Instrument data release 1 (DESI DR1), when combined with Cosmic Microwave Background (CMB) data from Planck and Supernovae of Type Ia (SN Ia), favor evolving dark energy over cosmological constant. This result is strongly related to the assumed priors on the Chevallier-Polarski-Linder (CPL) parameterization of DE. We test another parametrization which introduces two free parameters $n$ and $\alpha$, only $n$ is independent. Thus, it reduces the parameter space compared to the CPL model, which derives a more robust preference for evolving DE, if any. The model potentially produces three cosmological scenarios according to the values of its parameters. For $n=3$, the $\Lambda$CDM model is recovered, quintessence for $n<3$, and phantom for $n>3$. In the present study, we test the model on the background level, and, to our knowledge for the first time, on the linear perturbation level. Bayesian evidence analysis shows a weak preference ($\ln B \leq 1.8$) for dynamical DE in the phantom regime over the cosmological constant DE using Planck, DESI, and PantheonPlus \& SH0ES data, similarly the AIC analysis supports dynamical DE scenario for the same data. The model predicts current phantom DE $w_{de,0} = -1.073 \pm 0.032$ and $H_0=70.9\pm 1.4$ km/s/Mpc when Planck+DESI data is used, which decreases the tension with $H_0$ local measurements to $1.2\sigma$ level.
\end{abstract}

\section{Introduction}
\label{Sec:Intro}

After the discovery of cosmic accelerating expansion \cite{Riess1998,Perlmutter}, the field of cosmology flourished with the models proposed to explain this unexpected discovery. The simplest model suggested a new exotic component with a negative pressure called dark energy by introducing a constant, $\Lambda$, into Einstein's field equations in addition to Cold Dark Matter (CDM), that is $\Lambda$CDM model. It offers an excellent fit to the cosmological data obtained from different probes such as the CMB temperature, polarization power spectra, the BAO, and the Large-Scale Structure (LSS) data \cite{Planck18_params, BAO_6dF, SDSS_DR12, eBoss}. Nevertheless, the $\Lambda$CDM model faces some serious theoretical obstacles such as the large discrepancy between the predicted amount of dark energy and the sufficient quantity needed to derive the current accelerated expansion, known as the cosmological constant problem \cite{Weinberg1988, Carroll1992} and the still-undetermined nature of the dark matter and dark energy components. Moreover, it has observational challenges, such as the Hubble tension and the $S_8 \, (\equiv \sigma_8 \sqrt{\Omega_m/0.3})$ tension. The Hubble tension is a discrepancy between the direct measurement of the Hubble parameter $H_0$ and the predicted value from the early universe data assuming the $\Lambda$CDM model, which results in a $3 - 5 \sigma$ tension between these two approaches \cite{Riess2021, Chen2019, Pascale2024}. On the other hand, the $S_8$ tension is less severe, $\approx 2.3 \sigma$, but still points out towards some cracks in the $\Lambda$CDM model \cite{Joudaki2016}. Although it is not easy, at least from a practical point of view, to replace $\Lambda$CDM model, the model may not be the best choice to fit all the cosmological data at present \cite{Peracaula2016, Yang2021, Staicova2021}. For more details on the different challenges facing $\Lambda$CDM and the possible solutions, see \cite{Bull2015, Brax2018, Vagnozzi:2019ezj, Dutta2020, Perivolaropoulos2021, H0_Olympics,realm_of_Hubble, Abdalla2022,challenges_LCDM, Hu2023, Toda:2024ncp,Bamba_2012}.

The presented issues motivate the search for an alternative model that can tackle all or some of these challenges. There are two main approaches to reaching this desired model: the first assumes general relativity as the correct theory of gravity while modifying the energy-momentum tensor. Some examples are Chaplygin gas \cite{Bento2002, Gorini2004}, unified dark fluid model \cite{Elkhateeb:2017oqy,Elkhateeb:2018unf,Elkhateeb:2021atj}, emergent DE \cite{Pan2019, Li2019,GEDE20}, interacting DE \cite{IDE_early_20,Gariazzo2021,Zhai2023,Giare:2024smz}, holographic DE \cite{Fischler1998, Cardona2022}, quintessence \cite{Zlatev1999, Brax1999, Wang2000,Wolf_2024} and dynamical DE (DDE). For a more comprehensive review of dark energy and the various models, one can consult \cite{Copeland2006}. The second main approach assumes that the standard model of particles is sufficient, whereas the gravitational sector can be modified to source the current accelerated expansion. There are various ideas to modify the Einstein field equations, such as the scalar-tensor theories (e.g. Brans-Dicke theory \cite{Brans_Dicke}), $f(R)$ theories \cite{Barrow1983, Nojiri2003, Carroll2003, Nojiri2004, Felice2010}, higher-dimensional theories (e.g. Kaluza-Klein theory \cite{Bailint1987}). Notably, a complete spectrum of infrared corrections of gravity within $f(T)$ modified gravity, without introducing extra free parameters, has been suggested to describe late cosmic expansion \cite{Awad:2017yod,Hashim:2020sez, Hashim:2021pkq}. Several other modifications are also available, c.f. \cite{Clifton2011, Capozziello2011, Capozziello2019,ElHanafy:2015jbo,ElHanafy:2017xsm,Awad:2017yod,Nojiri_2017, ElHanafy:2020pek}. 

The DDE approach is interesting because it builds the dark energy component phenomenologically, rather than relying on a field theory, which is a theoretical hardship. Therefore, it is a step that enhances our understanding of dark energy and hence can be a useful guide for follow-up physical field theory. Nevertheless, a connection can be found between a DDE model and a modified gravity model \cite{Capozziello2005, Nojiri_2005, Dent2011, MWN}. The most famous parametrization of DDE is the CPL model \cite{Chevallier2000, Linder2002} where the equation of state (EoS) is a linear function of the scale factor, i.e. $w(a)=w_0+w_a(1-a)$, characterized by two parameters $w_0$ and $w_a$. Models with a linear EoS in redshift or cosmological time are also explored \cite{Akarsu2015}. These, and other models where a function form for the dark energy EoS is assumed, belong to the parametric dark energy models class. Since dark energy affects the background evolution of the universe through the Hubble function, parametric dark energy can be described not just through its EoS \cite{Barboza2009, Ma2011, Sello2013, Pantazis2016, Yang2019, Dahmani2023}, but through the Hubble function \cite{Pacif2016, Roy2022} or any other quantity that depends on it, such as the dark energy density \cite{Barboza2012, Eleonora2020}, the deceleration parameter $q(z)$ \cite{Shapiro2005, Gong2006, Xu2008, Koussour2023}, the jerk parameter $j(z)$ \cite{Rapetti2007, Mukherjee2016_2}, etc. One can review the appendix of \cite{Pacif2020} for some parametric dark energy models.

In general, DDE models possess three potential behaviors; quintessence DE ($-1/3>w_{de} > -1$), phantom DE ($w_{de} < -1$), or quintom DE where the dark energy EoS crosses the phantom divide line $w_{de} = -1$ \cite{Cai2010}. It has been shown that to simultaneously solve both the $H_0$ and $\sigma_8$ tensions the dark energy EoS must cross the phantom divide line if the gravitational constant is held constant \cite{Heisenberg2022}. A further study shows that a sharp transition at $z\lesssim 0.2$ in the absolute magnitude of supernovae of Type Ia must be accompanied when angular BAO data is employed \cite{Gomez-Valent:2023uof}. Furthermore, quintom behavior has been shown to be consistent with DESI DR1 observations \cite{Roy:2024kni,Yang:2024kdo}. In the present study, we re-investigate a particular parametric effective EoS given by Mukherjee which introduces two free parameters $n$ and $\alpha$, whereas only $n$ is an independent \cite{Mukherjee_MWN}. Therefore, it is a one-parameter model, reducing the free parameters by one, compared to CPL parametrization which has been considered as a baseline in DESI study \cite{DESI}. This feature is important to derive a more robust preference for DDE, if any. The present model can potentially produce three DE scenarios: cosmological constant ($n=3$), quintessence ($n<3$), and phantom ($n>3$) with no divergences all over the redshift range $-1<z<\infty$. The model was tested against background data only, and it was found that the dark energy component behaves as pure quintessence. The same model has been reconciled within $f(T)$ modified gravity framework on the background level \cite{MWN}. We note that the recent BAO measurements by DESI DR1 when combined to CMB from Planck and Pantheon SNIa data favors DDE over $\Lambda$CDM \cite{DESI}, see also \cite{Giare:2024gpk}. However, this conclusion strongly depends on the assumed priors on the CPL parametrization \cite{Cortes:2024lgw, gialamas2024interpretingdesi2024bao, dinda2024modelagnosticassessmentdarkenergy}. In this sense, other parametrizations, which potentially can produce several DE scenarios while reducing the parameter space as in Mukherjee parameterization, need to be carefully studied. It is the aim of the present work to confront the mentioned DDE model with the latest cosmological observations including full Planck data which has not been studied yet. 

This paper is organized as follows: in Section \ref{Sec:Model}, we introduce the Mukherjee model and the different aspects associated with its parameters $n$ and $\alpha$. Section \ref{Sec:Data} illustrates the methodology and datasets used to constrain the model parameters. We use \texttt{CLASS} to calculate the cosmological quantities predicted by the model and Markov chain Monte Carlo (MCMC) algorithm through \texttt{MontePython} to find the best-fit values for the model's parameters using CMB data from the Planck satellite, DESI BAO data, PantheonPlus SN sample and Supernovae and $H_0$ for the Equation of State of dark energy (SH0ES) data. Section \ref{Sec:Results} is devoted to results and discussion. Also, we discuss the significance of the model in comparison to $\Lambda$CDM using different data. We conclude the work in Section \ref{Sec:Summary} and expand more on the effect of the model on the local universe in the Appendix.

\section{The Model parameterization}
\label{Sec:Model}
In this section, we utilize a particular parameterization of an effective equation of state, which has been introduced in \cite{Mukherjee_MWN}, to describe the cosmological history. We derive the corresponding dark energy equation of state and its consequences on cosmic evolution on the background and the linear perturbation levels.

\subsection{Background Description}

We assume that the background spacetime of the universe is described by a flat Friedmann-Lemaître-Robertson-Walker (FLRW) metric\footnote{We use the units in which the speed of light $c=1$.} 
\begin{equation}
    ds^2 = -dt^2 + a^2(dx^2 + dy^2 + dz^2)
\end{equation}
where $a\equiv a(t)$ is the scale factor. By plugging this metric into the Einstein field equations, one obtains the Friedmann equations
\begin{align}
 3H^2(t) &= 8\pi G \sum_i\rho_i(t) = 8\pi G \rho_{eff}(t), \label{eq:FR1}\\
 2\dot{H}(t) + 3H^2(t) & = -8\pi G \sum_i P_i(t) = -8\pi G P_{eff}(t), \label{eq:FR2}
\end{align}
where $H=\dot{a}/a$ is the Hubble function, $G$ denotes Newtonian constant, $P_{eff}(t)$ is the total (effective) pressure of all components, $\rho_{eff}(t)$ is their total (effective) energy density. The index $i$ runs over the different species, matter, radiation, dark energy, \dots, and the overhead dot means derivative with respect to cosmic time $t$. The second Friedmann equation is usually replaced by the continuity equation
\begin{equation}
    \label{continuity}
    \dot{\rho}_{eff}(t) + 3 H(t) [1+w_{eff}(t)] \rho_{eff}(t) = 0,
\end{equation}
by imposing an effective equation of state for the compound fluid, i.e. $w_{eff}=P_{eff}/\rho_{eff}$. Notably, the continuity equation is also valid for any noninteracting component.

Given any of the functions $H(t), \rho_{eff}(t)$, or $ w_{eff}(t)$, one can solve the above system of equations and find how the scale factor evolves with time. There are other kinematical quantities, other than $H(t)$, such as the deceleration parameter $q(t)$, and the jerk parameter $j(t)$, which can be defined from the Taylor expansion of the scale factor $a(t)$ around $t = t_0$, where $t_0$ is the age of the universe and subscript "0" means evaluated at the present time.
\begin{eqnarray}
\nonumber    \frac{a(t)}{a(t_0)} &=&1 + \frac{\dot{a}}{ a}\biggr\rvert_{t=t_0}(t - t_0) + \frac{1}{2}\frac{\ddot{a}}{a}\biggr\rvert_{t=t_0}(t-t_0)^2 + \frac{1}{6}\frac{\dddot{a}}{a}\biggr\rvert_{t=t_0}(t-t_0)^3+\dots\\
     &=&1 + H_0(t - t_0) - \frac{1}{2} q_0 H^2_0(t-t_0)^2 - \frac{1}{6}j_0 H^3_0(t-t_0)^3+\dots
\end{eqnarray}
Hence, one can define the functions
\begin{equation}\label{Deceleration-Jerk Definition}
q(t) \equiv - \frac{\ddot{a}}{aH^2}, \text{ and } j(t) \equiv - \frac{\dddot{a}}{aH^3}.
\end{equation}
Therefore, by assuming a function form for any of these quantities, the scale factor can be obtained, up to a few constants that data can constrain.

For a flat FLRW spacetime, the Hubble function $H(z)$ can be written, in general, as
\begin{equation}
    \label{Genaral_Hubble}
    H^2(z) = H^2_0\left[\Omega_{m,0}(1+z)^3 + \Omega_{r,0}(1+z)^4 + X(z)\right],
\end{equation}
where the density parameter at present $\Omega_{i,0} \equiv \rho_{i,0}/\rho_{critical} = 8\pi G\rho_{i,0}/3H^2_0$, and $i = m, r, de$, respectively represent matter, radiation and dark energy. Thus, the dark energy density evolution can be described by the function
\begin{equation}\label{eq:X(z)}
    X(z) = \Omega_{de,0} \, \exp\left({\displaystyle{\int_0^z} \frac{3(1+w_{de}(\bar{z}))}{1+\bar{z}} d\bar{z}}\right),
\end{equation} 
where $\Omega_{de,0}= 1-\Omega_{m,0}-\Omega_{r,0}$, the EoS of the dark energy $w_{de}(z)=p_{de}/\rho_{de}$ is given as a function of the redshift $z= -1+1/a$.

In this work, we explore the following effective EoS \cite{Mukherjee_MWN} and its consequences on the cosmological evolution in the late universe (after matter-radiation equality\footnote{We use tilde above a parameter when the radiation contribution to the cosmic expansion is ignored, i.e. after matter-radiation equality.})
\begin{equation}
\label{w_eff}
    \widetilde{w}_{eff} = \frac{p_{m}+p_{de}}{\rho_{m}+\rho_{de}}=\frac{-1}{1+\alpha(1+z)^n},
\end{equation}
where $\alpha$ and $n$ are the model parameters. One can notice that the above parameterization combines both the cosmic evolution at matter domination and dark energy domination eras. At large $z$, the model recover the standard CDM scenario, i.e. $\widetilde{w}_{eff} \to 0$, while the universe evolves toward de Sitter as $z\to -1$ where $\widetilde{w}_{eff}\to -1$. In this sense, we can derive the corresponding Hubble function, $\widetilde{H}(z)$, during these epochs, since
\begin{equation}
   \widetilde{w}_{eff} =-1+\frac{2}{3}(1+z)\frac{d\ln{\widetilde{H}(z)}}{dz}.
\end{equation}
Consequently, we have
\begin{equation}
    \label{Hubble_tilde}
    \widetilde{H}^2(z)=H^2_0\left(\frac{1+\alpha(1+z)^n}{1+\alpha}\right)^\frac{3}{n},
\end{equation}
where radiation is negligible at late evolution. Obviously, for $n=3$, the model mimics $\Lambda$CDM where the model parameter $\alpha$ is completely determined by the current matter density parameter as $\alpha=\frac{\Omega_{m,0}}{1-\Omega_{m,0}}$, equivalently $\Omega_{m,0}=\frac{\alpha}{1+\alpha}$. Using Equation \eqref{Deceleration-Jerk Definition}, one obtains the deceleration $q(z)$ and the jerk $j(z)$ parameters corresponding to the parameterization \eqref{w_eff} at late universe \cite{Mukherjee_MWN}
\begin{eqnarray}
    \widetilde{q}(z) &=& -1 + \frac{3\alpha(1+z)^n}{2[1+\alpha(1+z)^n]},\\[5pt]
    \widetilde{j}(z) &=& -1 - \frac{3\alpha(n-3)(1+z)^n}{2[1+\alpha(1+z)^n)]} + \frac{3\alpha^2(n-3)(1+z)^{2n}}{2[1+\alpha(1+z)^n]^2}.
\end{eqnarray}

It is straightforward to obtain the dark energy density evolution
\begin{equation*}
    X(z)=\left(\frac{1+\alpha(1+z)^n}{1+\alpha}\right)^\frac{3}{n}-\Omega_{m,0}(1+z)^3.
\end{equation*}
Consequently, using Equation \eqref{eq:X(z)}, one can derive the EoS of the dark energy 
\begin{eqnarray}
    \label{w_de_modified}
\nonumber    w_{de} &=&-1+\frac{1}{3}(1+z) \frac{d \ln{X(z)}}{dz},\\
    &=& - \frac{[1+\alpha(1+z)^n]^{\frac{3}{n}} - \alpha (1+z)^n [1+\alpha(1+z)^n]^{\frac{3}{n}-1}}{[1+\alpha(1+z)^n]^{\frac{3}{n}} - \Omega_{m,0}(1+\alpha)^{\frac{3}{n}}(1+z)^3}.
\end{eqnarray}
To account for radiation contribution to the cosmic evolution, assuming no interaction between different components, we can solve the continuity equation, as given in Equation \eqref{continuity}, for each component. Thus, the Friedmann equation \eqref{eq:FR1} reads
\begin{equation}
    \label{Hubble_MWN}
    H^2(z)=H^2_0\left\{\left[\frac{1+\alpha(1+z)^n}{1+\alpha}\right]^\frac{3}{n} + \Omega_{r,0}(1+z)^4\right\},
\end{equation}
where the first term encapsulates the contribution of both components of the dark sector (dark matter and dark energy) while the second term describes the radiation contribution.

\begin{figure}
    \centering
    \includegraphics[width=0.48\textwidth]{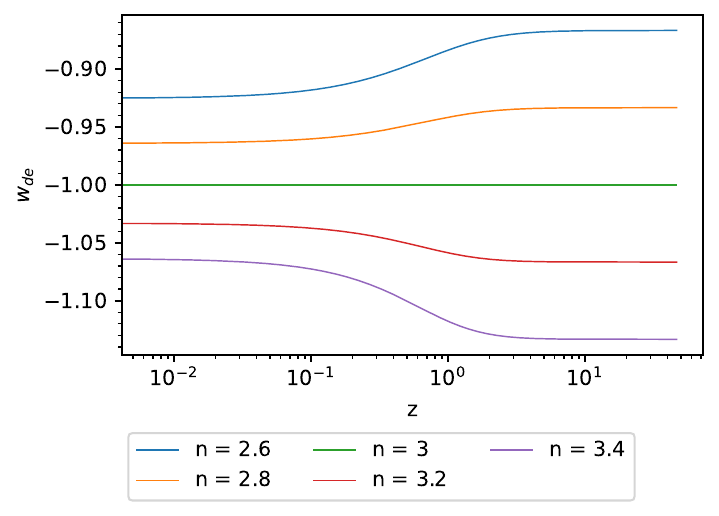}\hspace{5pt}
    \includegraphics[width=0.48\textwidth]{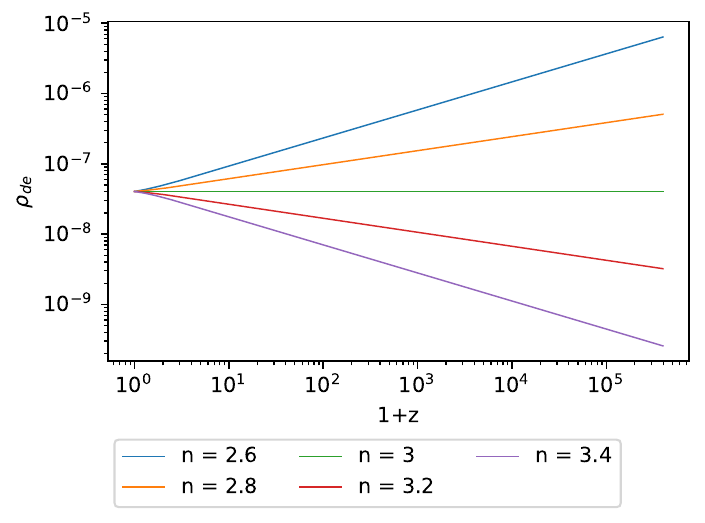}
    \caption{ \textit{Left}: Dark energy equation of state for various values of the model parameter $n$. \textit{Right}: Dark energy physical density for various values of the model parameter $n$. We assume that $\Omega_{m,0} = 0.3$ and $H_0 = 70$ km/s/Mpc. Dark energy density $\rho_{de}$ has units of $Mpc^{-2}$.}
    \label{Fig:Mukherjee_DE}
\end{figure}

From the series expansion of the Hubble function in Equation \eqref{Hubble_MWN}, $n\neq 3$, we obtain a relation between the matter density parameter $\Omega_{m,0}$ and the model parameters $n$ and $\alpha$ by identifying the term $(\frac{\alpha}{1+\alpha})^{3/n} (1+z)^3$ as the matter density $\Omega_{m,0} (1+z)^3$. This allows us to put a constraint on the model parameters, hence reducing the degrees of freedom,
\begin{equation}
    \label{Omega_m_constriant}
    \Omega_{m,0} = \left(\frac{\alpha}{1+\alpha}\right)^{3/n},\text{ equivalently } \alpha=\frac{\Omega_{m,0}^{n/3}}{1-\Omega_{m,0}^{n/3}}
\end{equation}
We choose to consider $n$ as a free parameter and fix $\alpha$ using the constraint (\ref{Omega_m_constriant}). In this sense, the dark energy EoS \eqref{w_de_modified} has three different behaviors depending on the value of $n$; for $n=3$, the model reduces to $\Lambda$CDM, and for $n<3 \; (n>3)$, it behaves as quintessence (phantom). For $n\neq3$, the universe evolves to de Sitter $w_{de} \to -1$ in the far future as $z\to -1$, whereas one can show that $w_{de} \to -n/3$ at very early phases $z\to\infty$. This shows that the parametrization \eqref{w_de_modified} does not diverge in the full redshift range. We visualize the evolution of the EoS and the physical density, $\rho_{de}(z)\equiv X(z)H_0^2$, of the dark energy sector for different values of $n$ fixing $\Omega_{m,0}=0.3$ and $H_0=70$ km/s/Mpc as seen in Figure \ref{Fig:Mukherjee_DE}.

We note that the recent BAO measurements by DESI DR1 when combined to CMB from Planck and Pantheon SNIa data favor evolving DE over the fixed cosmological constant DE scenario. In fact, the DESI conclusion strongly depends on the assumed priors on the CPL DE parametrization, $w(a)=w_0+w_a(1-a)$, which has two free parameters. Clearly, the present DDE model, \eqref{w_de_modified}, is a one-parameter model. In this sense, if a preference for evolving DE ($n\neq 3$) is found, it would be more robust in this model.
\subsection{Perturbation Level}
In the $\Lambda$CDM model the dark energy is represented by a cosmological constant, hence the dark energy density is not perturbed. On the other hand, for dynamical dark energy models, the dark energy density is allowed to be perturbed. The perturbed metric in the Newtonian gauge can be written as
\begin{equation}
    ds^2 = a^2(\eta)[(1+\phi)d\eta^2 - (1-\psi)d\Vec{x}^2],
\end{equation}
where $\phi$ and $\psi$ are two scalar potentials that depend on space and time and quantify the deviation from the homogeneous FLRW background, and $\eta$ is the conformal time related to the cosmological time $t$ by $dt = a(\eta)d\eta$.

For each non-interacting fluid, we can obtain the energy and momentum conservation equations through the Bianchi identity $\nabla_{\mu}T^{\mu}{_\nu} = 0$, where $\nabla_{\mu}$ represents the covariant derivative operator and $T{^\mu}{_\nu}$ is the energy-momentum tensor. On the first-order perturbation level, the conservation equations, in the $k$-space, have the form \cite{Mukhanov1992TheoryOC,Ma:1995ey,Malik:2008im}
\begin{equation}
    \label{energy_pert_cons}
    \delta^\prime_i = -(1+w_i)(\theta_i - 3\phi^\prime) - 3 \frac{a^\prime}{a}\left(\frac{\delta P_i}{\rho_i} - w_i \delta_i\right),
\end{equation}
\begin{equation}
    \label{velocity_pert_cons}
    \theta^\prime_i = \frac{a^\prime}{a}(3w_i-1)\theta_i - \frac{w^\prime_i}{1+w_i}\theta_i + k^2\left(\frac{\delta P_i/\rho_i}{1+w_i} + \phi)\right),
\end{equation}
where $\delta_i \equiv \delta\rho_i/\rho_i$ and $(\rho_i+P_i)\theta_i \equiv \partial^{j} \delta T^0_{(i)j} $ are the energy density contrast and velocity potential perturbations of the $i$th species respectively, while $(^\prime)$ denotes derivative with respect to conformal time.

The pressure perturbation $\delta P_i$ is given by
\begin{equation}
    \label{pressure_pert}
    \delta P_i = c^2_{s,i} \delta\rho_i + 3\frac{a^\prime}{a}(1+w_i)\frac{\rho_i}{k^2}\left(c^2_{s,i} - c^2_{a,i}\right)\theta_i,
\end{equation}
where $c^2_{a,i}$ and $c^2_{s,i} \equiv \frac{\delta P^{(rest)}_i}{\delta \rho^{(rest)}_i}$ are the adiabatic and rest-frame sound speeds, respectively.

From Equation (\ref{velocity_pert_cons}), we can see that the velocity perturbation diverges if $w_i = -1$ and $c^2_{s,i}$ is held fixed, which leads to gravitational instability. This requires a different approach to find $\theta_i$ other than the pressure perturbation. To allow for the cosmological constant case ($n=3$) and dynamical behavior for the dark energy EoS ($n\neq 3$), we use the Parameterized Post-Friedmann (PPF) approximation \cite{PPF_genearl, PPF_phantom}. The PPF approximation replaces the relation between the dark energy density perturbation $\delta\rho_{de}$ and pressure perturbation $\delta P_{de}$ with a relation between the dark energy velocity perturbation $\theta_{de}$ and that of the rest of the fluids on large scales. In this approximation, a new dynamical parameter $\Gamma$ is defined and related to the dark energy density perturbation in the dark energy rest frame via
\begin{equation}
    \Gamma = -\frac{4\pi Ga^2}{k^2 c_K}\delta\rho_{de}^{rest},
\end{equation}
where $c_K = 1-3K/k^2$, and $K$ is the background curvature. The evolution equation for $\Gamma$ is
\begin{equation}
    \left(1+\frac{c^2_{\Gamma} k^2}{a^2H^2}\right)\left[\frac{\Gamma^\prime}{aH}+\left(1+\frac{c^2_{\Gamma} k^2}{a^2H^2}\right)\Gamma\right] = S,
\end{equation}
where $c_\Gamma = 0.4 c_{s, de}$\cite{PPF_phantom}. The source term $S$ is given by
\begin{equation}
    S = \frac{a^\prime}{a}\frac{4\pi G}{H^2}\rho_{de}(1+w_{de})\frac{\theta_T}{k^2},
\end{equation}
where $\theta_T$ is the total perturbed velocity divergence of all species except dark energy.

Solving the differential equation of $\Gamma$ allows one to find the dark energy density perturbation in the rest frame. In any other gauge, the dark energy density perturbation can be found by
\begin{equation}
    \delta\rho_{de} = \delta\rho^{rest}_{de} - 3\frac{a^\prime}{a}\rho_{de}(1+w_{de})\frac{\theta_{de}}{k^2}
\end{equation}
The above equations behave nicely when $w_{de} = -1$, hence, the PPF approximation offers the evolution of the perturbed quantities smoothly when the DE has a dynamical evolution or just a cosmological constant.

\section{Data and Methodology}
\label{Sec:Data}

The parameter space for this model consists of the regular six $\Lambda$CDM parameters; the physical baryon energy density $\Omega_bh^2$, where $h=H_0/(100~\text{km/s/Mpc})$ denotes the dimensionless Hubble parameter, the physical cold matter energy density $\Omega_ch^2$, the angular scale of the sound horizon at decoupling $\theta_s$, the amplitude and spectral index of the primordial scalar spectrum $A_s, n_s$, and the reionization optical depth $\tau_{reio}$, which we will hereafter assume the \textit{base} parameters, in addition to the model parameter $n$. To constrain the parameters, we consider the following datasets:
\begin{itemize}
    \item \textit{Planck}: CMB temperature anisotropy and polarization power spectra by Planck 2018 and their cross-correlations in addition to lensing power spectrum:
    \begin{description}
        \item [(i)]  We use the \textit{plik} Likelihood \textit{Planck high\_l\_TTTEEE\_lite} along the \textit{Planck\_lowl\_EE} and \textit{Planck\_lowl\_TT} likelihoods \cite{Planck18_likelihoods}.
        \item [(ii)] CMB lensing reconstruction power spectrum obtained from the CMB trispectrum analysis by Planck 2018 \cite{Planck18_lensing}.
    \end{description}
    \item We use the recently published Dark Energy Spectroscopic Instrument (DESI) Year 1, which includes the transverse comoving distance, the Hubble horizon, and the angle-averaged distance \cite{DESI}. We referred to it as \textit{DESI}. \footnote{During the final stage of this work, DR2 have been released. However,  We do not expect any qualitative changes when replacing DESI DR1 by DR2.}
    \item Supernovae Type Ia (SNIa) distance moduli data from the PantheonPlus sample \cite{Scolnic:2021amr,Brout:2022vxf}, referred to as \textit{PantheonPlus} or as \textit{SN} in some text. This includes 1550 spectroscopic SNIa in the range $0.001 < z < 2.26$. However, we remove low-redshift data points $z<0.01$ which are sensitive to the peculiar velocities of the host galaxies \cite{Peterson:2021hel}. 
    \item Supernovae Type Ia (SNIa) distance moduli data from the PantheonPlus sample \cite{Scolnic:2021amr,Brout:2022vxf} and SH0ES \cite{Riess2021}, referred to as \textit{PantheonPlus\&SH0ES}. This is to include local universe measurements by adding prior on $H_0$ and consequently the absolute magnitude $M$.
\end{itemize}
We note that a prior from Big Bang Nucleosynthesis (BBN) on the baryon energy density $\omega_b=\Omega_b h^2$ \cite{BBN_class} is being used with all the above datasets. For the radiation component, $\Omega_{r,0}=\Omega_{\gamma,0}\left[1+\frac{7}{8}\left(\frac{4}{11}\right)^{4/3} N_{eff}\right]$, where $\Omega_{\gamma,0}=2.4728\times 10^{-5} h^{-2}$,
we set the effective number of neutrinos to the Standard Model value $N_{eff} \approx 3.044$, and the average CMB temperature $T_0 = 2.7255$ \cite{Planck18_params}. We use Cosmic Linear Anisotropy System Solver (\texttt{CLASS}) \cite{CLASS} to calculate the various cosmological quantities for this model and $\Lambda$CDM. We sample the posterior distribution of the cosmological parameters by making use of the Markov Chain Monte Carlo (MCMC) Metropolis-Hastings algorithm using the publicly available \texttt{MontePython} \cite{MontePython}, where all chains fulfill the Gelman-Rubin convergence criterion $R-1 < 0.01$ \cite{Rubin_Gelman} for all parameters. All contour plots and tables presented in this paper are obtained using \texttt{GetDist} \cite{getDist}. The assumed priors are shown in Table \ref{table:priors}.
\begin{table}
    \centering
   \caption{Flat priors on the cosmological parameters used in this work.}
    \begin{tabular}{ll}
    \noalign{\vskip 3pt}\hline \hline
    Parameter & Prior \\[3pt]
    \hline \hline\noalign{\vskip 5pt}
$\Omega_b h^2$  & $\mathcal{U}$[0.005, 0.03] \\[3pt]
$\Omega_c h^2$  & $\mathcal{U}$[0.01, 0.2] \\[3pt]
$100\theta_s$  & $\mathcal{U}$[0.5, 2] \\[3pt]
$\ln (10^{10}A_s)$  & $\mathcal{U}$[2, 4] \\[3pt]
$n_s$  & $\mathcal{U}$[0.7, 1.2] \\[3pt]
$\tau_{reio}$  & $\mathcal{U}$[$4\times10^{-3}$, 0.1] \\[3pt]
$n$  & $\mathcal{U}$[2, 5] \tablefootnote{We expect the bestfit value $n\sim3$, so we choose a flat prior around $3$. The prior range is asymmetric because the CMB-alone dataset requires a value of $n>4$, as will be seen in Table \ref{tab:MWN}; hence, we extend the upper end of the prior range.} \\
\hline
    \end{tabular}
    \label{table:priors}
\end{table}

\section{Results and Discussion}
\label{Sec:Results}
In Table \ref{tab:LCDM}, We show the parameters' best-fit values with the 68\% constraint for $\Lambda$CDM for different combinations of the datasets mentioned in the previous section. In Table \ref{tab:MWN}, we present the best fit of the parameters and the constraints on these parameters at 68\% CL for the proposed model using the same combinations of datasets\footnote{We note that all parameters have been reproduced using \texttt{Cobaya} sampling code \cite{Torrado:2020dgo} with CMB lensing likelihoods from Planck and Atacama Cosmology Telescope (ACT), DESI, PantheonPlus and PantheonPlus\&SH0ES. Only negligible shifts in the parameters are noticed, whereas all conclusions presented in this study remain the same.}. Both tables are divided into three parts: The first is for the cosmological parameters which are the \textit{base} parameters, in the case of $\Lambda$CDM, in addition to the parameter $n$, in the case of the Mukherjee model. The second part contains the apparent magnitude $M$ as a nuisance parameter for the PantheonPlus and PantheonPlus\&SH0ES datasets. The third part gives the derived parameters which can be listed as follows; the present-day matter $\Omega_{m,0}$ and dark energy $\Omega_{de,0}$ density parameters, the Hubble constant $H_0$, the normalization of the linear matter power spectrum $\sigma_8$ and the related quantity $S_8$, the age of the universe, the sound horizon at recombination $r_{s,rec}$ with the corresponding redshift $z_{rec}$, the pivot scale at matter-radiation equality $k_{eq}$ with the corresponding redshift $z_{eq}$, the sound horizon at baryon drag $r_{d}$, the dark energy equation of state $w_0$, the deceleration parameter $q_0$, and the jerk parameter $j_0$.

\begin{table}
    \centering
    \caption{68\% CL parameters for $\Lambda$CDM from Planck data, DESI, PantheonPlus, and SH0ES. The first six parameters are the \textit{base} parameters.}
\begin{adjustbox}{width=\textwidth}
\begin{tabular} { l  c c c c}
\noalign{\vskip 3pt}\hline\noalign{\vskip 1.5pt}\hline\noalign{\vskip 5pt}
\multicolumn{1}{c}{Parameter} &
\multicolumn{1}{c}{Planck} &
\multicolumn{1}{c}{\makecell{~~Planck\\+ DESI}} &
\multicolumn{1}{c}{\makecell{~~Planck + DESI \\+ PantheonPlus}} &
\multicolumn{1}{c}{\makecell{Planck + DESI \\~~+ PantheonPlus\&SH0ES}} \\
\noalign{\vskip 3pt}\hline\noalign{\vskip 1.5pt}\hline\noalign{\vskip 5pt}
\vspace{5pt}
{$\Omega_bh^2    $} & $0.02235\pm 0.00014$ & $0.02250\pm 0.00014$ & $0.02247\pm 0.00013$ & $0.02265\pm 0.00013$\\
\vspace{5pt}
{$\Omega_ch^2    $} & $0.1203\pm 0.0012  $ & $0.11832\pm 0.00089$ & $0.11871\pm 0.00085$ & $0.11700\pm 0.00079$\\
\vspace{5pt}
{$100\theta{}_{s}$} & $1.04185\pm 0.00029$ & $1.04202\pm 0.00028$ & $1.04200\pm 0.00028$ & $1.04220\pm 0.00028$\\
\vspace{5pt}
{$\ln(10^{10}A_{s})$} & $3.047^{+0.013}_{-0.015}   $ & $3.051\pm 0.015$ & $3.050\pm 0.015$ & $3.059\pm 0.016$\\
\vspace{5pt}
{$n_{s}  $} & $0.9645\pm 0.0041  $ & $0.9696\pm 0.0036  $ & $0.9686\pm 0.0036  $ & $0.9730\pm 0.0035  $\\
\vspace{5pt}
{$\tau_{reio}    $} & $0.0553^{+0.0067}_{-0.0078}$ & $0.0591^{+0.0068}_{-0.0076}$ & $0.0579\pm 0.0074  $ & $0.0640^{+0.0073}_{-0.0084}$\\
\noalign{\vskip 2pt}\noalign{\vskip 2pt}
\hline
\noalign{\vskip 2pt}\noalign{\vskip 2pt}
{$M  $} & -- & -- & $-19.423\pm 0.012  $ & $-19.398\pm 0.010  $\\
\noalign{\vskip 2pt}\noalign{\vskip 2pt}
\hline
\noalign{\vskip 2pt}\noalign{\vskip 2pt}
\vspace{5pt}
{$\Omega_{m,0}   $} & $0.3168^{+0.0070}_{-0.0079}$ & $0.3046\pm 0.0053  $ & $0.3070\pm 0.0051  $ & $0.2962\pm 0.0045  $\\
\vspace{5pt}
{$\Omega_{\Lambda,0}$} & $0.6832^{+0.0079}_{-0.0070}$ & $0.6953\pm 0.0053  $ & $0.6930\pm 0.0051  $ & $0.7037\pm 0.0045  $\\
\vspace{5pt}
{$H_0$} & $67.26\pm 0.54 $ & $68.15\pm 0.41 $ & $67.98\pm 0.39 $ & $68.83\pm 0.36 $\\
\vspace{5pt}
{$\sigma_8 $} & $0.8124^{+0.0054}_{-0.0061}$ & $0.8090\pm 0.0061  $ & $0.8095\pm 0.0060  $ & $0.8081\pm 0.0064  $\\
\vspace{5pt}
{$Age$} & $13.799\pm 0.023   $ & $13.769\pm 0.020   $ & $13.775\pm 0.019   $ & $13.741\pm 0.018   $\\
\vspace{5pt}
{$r_{s, rec}     $} & $144.47\pm 0.27$ & $144.86\pm 0.21$ & $144.78\pm 0.20$ & $145.08\pm 0.20$\\
\vspace{5pt}
{$z_{rec}$} & $1088.92\pm 0.22   $ & $1088.61\pm 0.18   $ & $1088.67\pm 0.18   $ & $1088.35\pm 0.17   $\\
\vspace{5pt}
{$100k_{eq} $} & $1.0405\pm 0.0083$ & $1.0273\pm 0.0061$ & $1.0299\pm 0.0059$ & $1.0188\pm 0.0055$\\
\vspace{5pt}
{$z_{eq} $} & $3409\pm 27    $ & $3366\pm 20    $ & $3374\pm 19    $ & $3338\pm 18    $\\
\vspace{5pt}
{$S_8$} & $0.835\pm 0.013$ & $0.815\pm 0.010$ & $0.8189\pm 0.0099  $ & $0.8030\pm 0.0094  $\\
\vspace{5pt}
{$r_{s, d} $} & $147.04\pm 0.27$ & $147.39\pm 0.22$ & $147.33\pm 0.22$ & $147.57\pm 0.21$\\
\vspace{5pt}
{$q_0$} & $-0.525^{+0.010}_{-0.012}  $ & $-0.5430\pm 0.0080 $ & $-0.5395\pm 0.0076 $ & $-0.5556\pm 0.0068 $\\
\hline
\end{tabular}
\end{adjustbox}
    \label{tab:LCDM}
\end{table}
\begin{table}
    \centering
    \caption{68\% CL parameters for the Mukherjee model (the $\Lambda$CDM \textit{base} parameters plus the model free parameter $n$) using Planck data, DESI, PantheonPlus, and SH0ES.}
\begin{adjustbox}{width=\textwidth}
\begin{tabular} {l  c c c c}
\noalign{\vskip 3pt}\hline\noalign{\vskip 1.5pt}\hline\noalign{\vskip 5pt}
\multicolumn{1}{c}{Parameter} &
\multicolumn{1}{c}{Planck} &
\multicolumn{1}{c}{\makecell{~~Planck\\+ DESI}} &
\multicolumn{1}{c}{\makecell{~~Planck + DESI \\+ PantheonPlus}} &
\multicolumn{1}{c}{\makecell{Planck + DESI \\~~+ PantheonPlus\&SH0ES}} \\
\noalign{\vskip 3pt}\hline\noalign{\vskip 1.5pt}\hline\noalign{\vskip 5pt}
\vspace{5pt}
{$\Omega_bh^2$} & $0.02239\pm 0.00015$ & $0.02240\pm 0.00014$ & $0.02246\pm 0.00014$ & $0.02251\pm 0.00014$\\
\vspace{5pt}
{$\Omega_ch^2$} & $0.1197\pm 0.0012$ & $0.1196\pm 0.0010$ & $0.11876\pm 0.00099$ & $0.11862\pm 0.00096$\\
\vspace{5pt}
{$100\theta{}_{s}$} & $1.04189\pm 0.00029$ & $1.04193\pm 0.00029$ & $1.04200\pm 0.00029$ & $1.04206\pm 0.00027$\\
\vspace{5pt}
{$\ln (10^{10}A_{s})$} & $3.042\pm 0.014$ & $3.046\pm 0.015$ & $3.052\pm 0.015$ & $3.051\pm 0.015$\\
\vspace{5pt}
{$n_{s}$} & $0.9660\pm 0.0041$ & $0.9663\pm 0.0038$ & $0.9684\pm 0.0038$ & $0.9688\pm 0.0038$\\
\vspace{5pt}
{$\tau_{reio}$} & $0.0535\pm 0.0073$ & $0.0553^{+0.0068}_{-0.0076}$ & $0.0589\pm 0.0078$ & $0.0587^{+0.0070}_{-0.0080}$\\
\vspace{5pt}
{$n$} & $4.16^{+0.81}_{-0.27}$ & $3.51^{+0.22}_{-0.28}$ & $3.00\pm 0.11$ & $3.30\pm 0.11$\\
\noalign{\vskip 2pt}\noalign{\vskip 2pt}
\hline
\noalign{\vskip 2pt}\noalign{\vskip 2pt}
{$M$} & -- & -- & $-19.425\pm 0.018$ & $-19.370\pm 0.015$\\
\noalign{\vskip 2pt}\noalign{\vskip 2pt}
\hline
\noalign{\vskip 2pt}\noalign{\vskip 2pt}
\vspace{5pt}
{$\Omega_{m,0}$} & $0.261^{+0.010}_{-0.031}   $ & $0.284\pm 0.011$ & $0.3077\pm 0.0067$ & $0.2882\pm 0.0054$\\
\vspace{5pt}
{$\Omega_{de,0}$} & $0.739^{+0.031}_{-0.010}$ & $0.716\pm 0.011$ & $0.6922\pm 0.0067$ & $0.7117\pm 0.0054$\\
\vspace{5pt}
{$H_0$} & $74.3^{+4.3}_{-1.6}$ & $70.9\pm 1.4$ & $67.92\pm 0.71$ & $70.14\pm 0.61$\\
\vspace{5pt}
{$\sigma_8$} & $0.871^{+0.036}_{-0.013}$ & $0.841\pm 0.016$ & $0.809\pm 0.010$ & $0.8288\pm 0.0096$\\
\vspace{5pt}
{$Age$} & $13.652^{+0.037}_{-0.088}$ & $13.715\pm 0.032$ & $13.777\pm 0.023$ & $13.721\pm 0.019$\\
\vspace{5pt}
{$r_{s, rec}$} & $144.58\pm 0.27$ & $144.61\pm 0.23$ & $144.77\pm 0.22$ & $144.77\pm 0.23$\\
\vspace{5pt}
{$z_{rec}$} & $1088.83\pm 0.22$ & $1088.81\pm 0.20$ & $1088.68\pm 0.19$ & $1088.62\pm 0.19$\\
\vspace{5pt}
{$100k_{eq}$} & $1.0369\pm 0.0085$ & $1.0357\pm 0.0070$ & $1.0303\pm 0.0068$ & $1.0295\pm 0.0066$\\
\vspace{5pt}
{$z_{eq}$} & $3397\pm 28$ & $3394\pm 23$ & $3376\pm 22$ & $3373\pm 22$\\
\vspace{5pt}
{$S_8$} & $0.810^{+0.015}_{-0.018}$ & $0.818\pm 0.010$ & $0.820\pm 0.010$ & $0.812\pm 0.010$\\
\vspace{5pt}
{$r_{s, d}$} & $147.14\pm 0.27$ & $147.17\pm 0.24$ & $147.31\pm 0.23$ & $147.30\pm 0.24$\\
\vspace{5pt}
{$w_0$} & $-1.129^{+0.011}_{-0.061}$ & $-1.073\pm 0.032$ & $-0.999\pm 0.020$ & $-1.047\pm 0.016$\\
\vspace{5pt}
{$q_0$} & $-0.755^{+0.025}_{-0.12}$ & $-0.653\pm 0.051$ & $-0.537\pm 0.028$ & $-0.618\pm 0.023$\\
\vspace{5pt}
{$j_0$} & $-1.1920^{-0.0053}_{-0.065}$&$-1.128^{+0.046}_{-0.059}$ & $-0.998\pm 0.036$ & $-1.084\pm 0.029$\\
\hline
\end{tabular}
\end{adjustbox}
\label{tab:MWN}
\end{table}

\begin{figure}
 \centering
 \includegraphics[width=\textwidth]{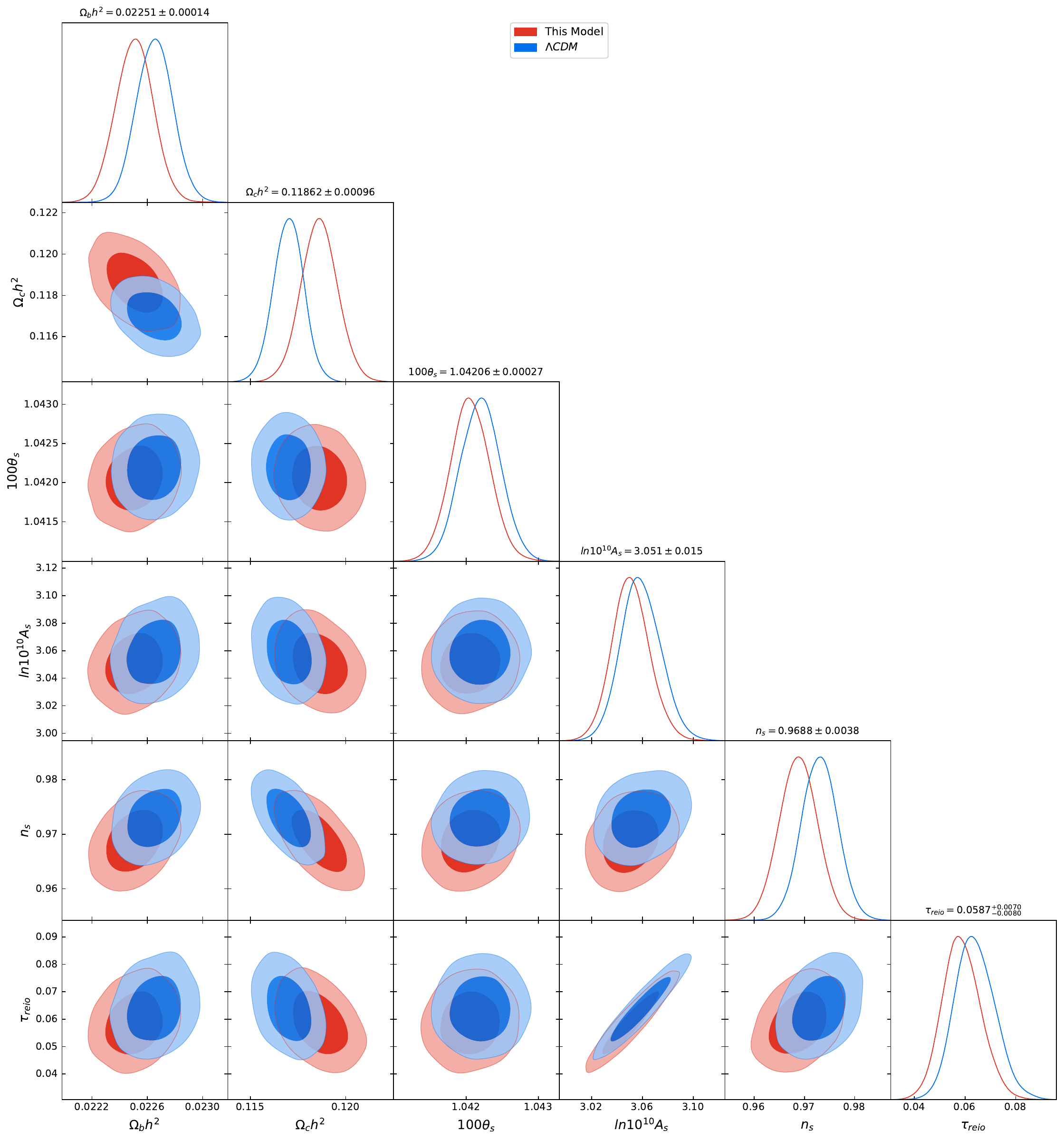}
    \caption{The 68\% (dark color) and 95\% (faint color) CL contours of the parameters for the $\Lambda$CDM and the Mukherjee model using the combination of \textit{Planck}, \textit{DESI} and \textit{PantheonPlus\& SH0ES} datasets. We also included the one-dimensional posterior distributions for the parameters.}
    \label{fig:contours}
\end{figure}

One can find that the best-fit values in Table \ref{tab:LCDM} agree with Planck results for $\Lambda$CDM model \cite{Planck18_params}. On the other hand, Table \ref{tab:MWN} shows that the base parameters of the present model, common with $\Lambda$CDM model, are in agreement within $1 \sigma$ for the different combinations of datasets, see also Figure \ref{fig:contours}. Notably, the Mukherjee model does not affect the inflationary phase, hence the initial conditions are similar to $\Lambda$CDM as can be realized via the agreement of the values of the two parameters $A_s$ and $n_s$ in both models. For the nuisance parameter $M$ of PantheonPlus\&SH0ES data, the absolute magnitude mean value in the Mukherjee model is relatively larger in comparison to its corresponding value in the $\Lambda$CDM model, which might explain the slight increase in the Hubble constant $H_0$ compared to $\Lambda$CDM. On the contrary, when PantheonPlus is used, both models have comparable values of $M$ and consequently $H_0$. However, the values of the derived parameters in both models vary according to the dataset used.

Remarkably, by considering the \textit{Planck} dataset alone in Tables \ref{tab:LCDM} and \ref{tab:MWN}, one finds that the parameters $\Omega_{m,0}$, $H_0$, and $\sigma_8$, as derived from the Mukherjee model, are at large deviations ($>4\sigma$) from those derived by the $\Lambda$CDM model. This can be attributed to the relatively large mean value of the model parameter $n=4.16^{+0.81}_{-0.27}$ indicating a deeper phantom regime. However, the \textit{Planck} dataset alone cannot set a strict constraint on the parameter $n$ as can be noticed. This is reasonable since the model parameters control the evolution of DE in the late universe, which manifests itself in the late integrated Sachs-Wolfe (ISW) contaminated by the large error bars due to the cosmic variance. We would like to stress that the results of \textit{Planck} dataset alone are prior-dependent and hence should not be interpreted as a physical behavior of the model. By adding late universe data (DESI, PantheonPlus, and SH0ES) to \textit{Planck} dataset, the deviations in the three parameters $\{\Omega_{m,0}, H_0, \sigma_8\}$ between the two models are getting small. Adding the \textit{DESI} dataset significantly constrains the model parameters, reducing the best-fit value of $n$ and the error bars to be $n=3.51^{+0.22}_{-0.28}$. Including \textit{PantheonPlus} puts a better constraint on $n = 3.00\pm 0.11$ which fixes the mean value of the model parameter at $\Lambda$CDM. However, by including local universe data as measured by \textit{SH0ES}, it gives $n = 3.30\pm 0.11$, which favors phantom DE over $\Lambda$CDM at the $2.7\sigma$ significance level. We discuss the consequences of the present DE model on the local cosmological parameters in Appendix \ref{app1}. As can be noticed, each combination prefers the phantom DE scenario except for Planck+DESI+PantheonPlus which clearly supports $\Lambda$CDM. We relate this result to the removal of the local universe data, $z<0.01$, in the PantheonPlus sample. We discuss this effect with the Bayesian model selection analysis in the next section.

\begin{figure}
    \centering
    \includegraphics[width=0.55\textwidth]{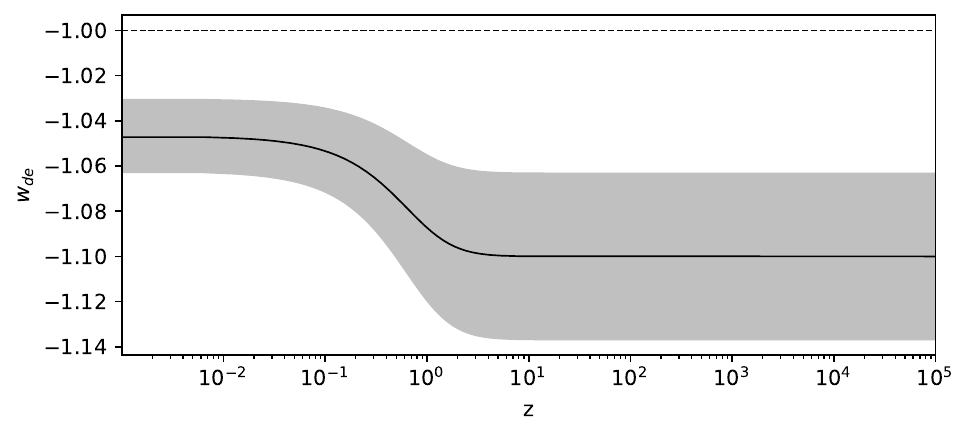}
    \caption{Evolution of the dark energy EoS with redshift using the best fit obtained using the combination of all datasets. The gray band represents the $1\sigma$ confidence region.}
    \label{fig:w_DE vs z}
\end{figure}

Referring to Table \ref{tab:MWN}, one finds $n>3$ at $1\sigma$ CL, indicating a phantom DE behavior with a current EoS parameter $w_0 = -1.047 \pm 0.016$ (68\% CL, \textit{All}). In Figure \ref{fig:w_DE vs z}, we plot the dark energy EoS evolution according to the obtained best-fit parameters. Our result differs from that has been obtained using background data only where the DE tends to be quintessence as $n = 2.907 \pm 0.136$ where the cosmological constant is not excluded at 1$\sigma$ CL \cite{Mukherjee_MWN}. Thus, the present study shows a tendency towards DDE in the phantom regime, $n=3.30\pm 0.11$ (All data), whereas the cosmological constant scenario is excluded at $1\sigma$ level. However, \textit{Planck} + \textit{DESI} + \textit{PantheonPlus} dataset gives $n=3.00\pm 0.11$, consequently the model shows no deviation from the cosmological constant DE, fixed to $w_{de}=-1$, at 1$\sigma$ at all redshifts.

There are several theoretical pathologies associated with phantom dark energy. If dark energy is modeled as a perfect fluid, it would violate the null energy condition. This includes the case of the canonical scalar field, minimally coupled to gravity, to model the dark energy. Phantom fields in this case exhibit a negative kinetic term, which triggers runaway particle production, hence violates unitarity and destabilizes quantum vacuum \cite{Nojiri_2005_03, Nojiri_2025}. Another issue with phantom fields is the exponential growth of small perturbations due to imaginary sound speed ``superluminal sound speed problem", causing catastrophic growth of anisotropies \cite{PPF_phantom}. Additionally, phantom fluids exhibit negative temperature and entropy decrease, violating the second law of thermodynamics \cite{K_M_2024_03,Brevik_2024_04}. For more details, see the review \cite{Ludwick:2017tox} (also \cite{Carroll:2003st,Carroll:2004hc,Ludwick:2015dba}). Notably in some scenarios one may allow phantom phase without ghost or gradient instabilities, if one extends $k$-essence to kinetic gravity braiding in which scalar stress tensor deviates from the perfect-fluid form \cite{Deffayet:2010qz}. Also, infrared corrections of gravity can mimic phantom dark energy. Since these corrections have a geometric origin not sourced by physical fluid, they do not violate physical laws, see \cite{Hashim:2020sez,Hashim:2021pkq, El-Zant:2018bsc}

\begin{figure}
    \centering
    \includegraphics[width=0.6\textwidth]{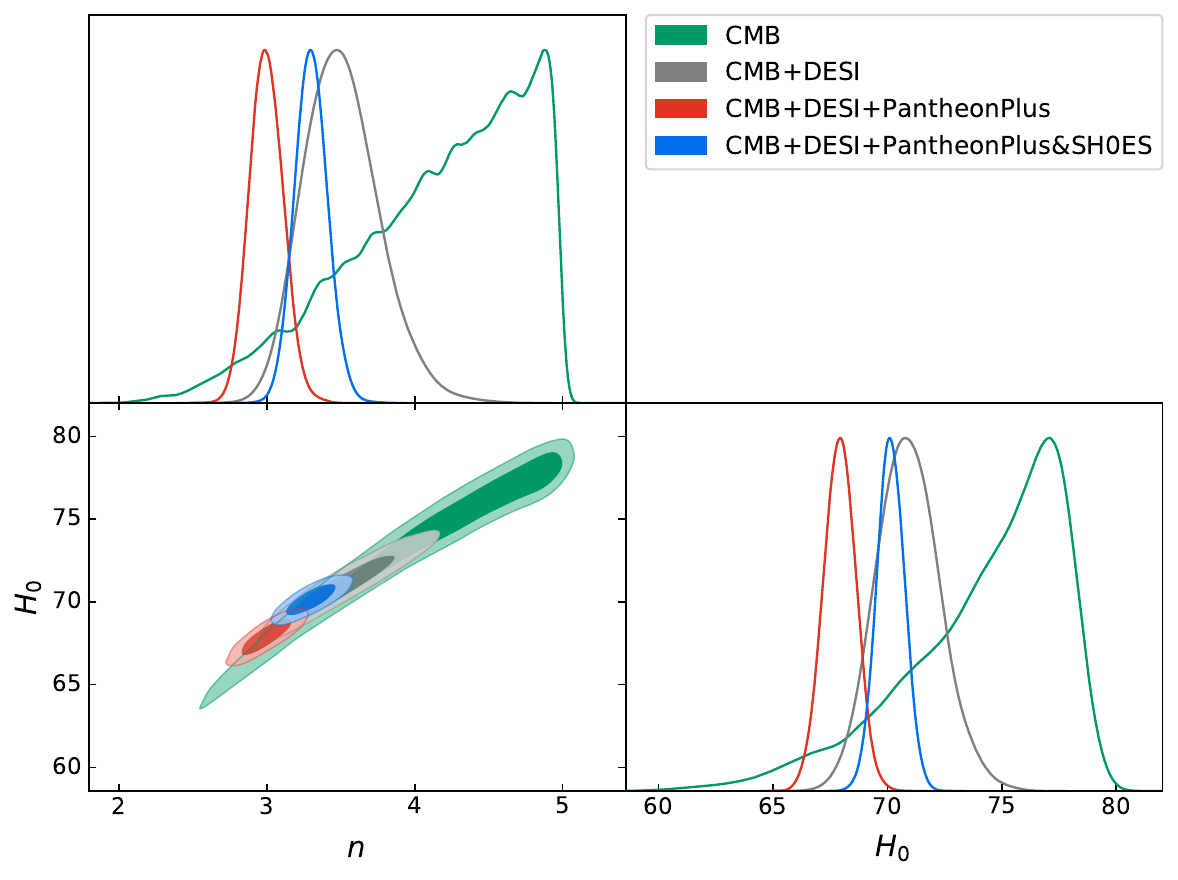}
        \caption{The correlation between $H_0$ and the model parameter $n$ for different dataset combinations. Clearly, Planck data alone cannot set a strict constraint, while late-time observations break the $n$-$H_0$ degeneracy and set better constraints.}
    \label{fig:H0_different_datasets}
\end{figure}

Figure \ref{fig:H0_different_datasets} shows the $n$-$H_0$ degeneracy, where a positive correlation between the two parameters can be realized. This is expected since higher $H_0$ values are associated with deeper phantom dark energy EoS, i.e. $n>3$, see \cite{El-Zant:2018bsc}. As mentioned above, \textit{Planck} dataset alone cannot set a strict constraint on $n=4.16^{+0.81}_{-0.27}$ as well as $H_0=74.3^{+4.3}_{-1.6}$  km/s/Mpc, while the parameter space is getting more confined when late-time data is included. The combination \textit{Planck} + \textit{DESI} gives $n=3.51^{+0.22}_{-0.28}$ and $H_0 = 70.9\pm 1.4$ km/s/Mpc, which decreases the tension with $H_0$ local measurements ($H_0=73.04\pm 1.04$ km/s/Mpc \cite{Riess2021}) to $1.2\sigma$, while $\Lambda$CDM is at $4.4\sigma$ tension and CPL parametrization is at $3.7\sigma$ level \cite{DESI} using same data. The combination \textit{Planck} + \textit{DESI} + \textit{PantheonPlus} gives more strict value $n=3.00 \pm 0.11$ and $H_0 = 67.92 \pm 0.71$ km/s/Mpc which reproduces $\Lambda$CDM scenario. We note that the lower $H_0$ value when using \textit{Planck} + \textit{DESI} + \textit{PantheonPlus} is consistent with lower $n$ value, but in tension with the other datasets. We relate this result to the removal of local data points, $z<0.01$, in the PantheonPlus sample which changes the absolute magnitude significantly. This will be discussed in the next section. Finally, the combination \textit{Planck} + \textit{DESI} + \textit{PantheonPlus}\&SH0ES provides the tightest constraint $n=3.30 \pm 0.11$ and $H_0 = 70.14 \pm 0.61$ km/s/Mpc, which prefers DDE phantom regime consistent with \textit{Planck} and \textit{Planck} + \textit{DESI} data.

It is to be mentioned that the true source of tension between late and early universe can be interpreted in term of the absolute magnitude of supernovae of Type Ia \cite{Camarena:2019moy,Benevento:2020fev,Camarena:2021jlr}. Indeed, the tension in the absolute magnitude between SH0ES measurement ($M=-19.253\pm 0.027$) and our result ($M=-19.370 \pm 0.015$ using all data) is $3.8\sigma$. But this is still better result than $\Lambda$CDM ($M=-19.398 \pm 0.010$) which raises the tension to $5\sigma$. On the other hand, SH0ES measurement determines $H_0= 73.04 \pm 1.04$ km/s/Mpc which is at 3.8$\sigma$ tension level with $\Lambda$CDM ($H_0 = 68.83 \pm 0.36$ km/s/Mpc) and 2.4$\sigma$ with Mukherjee parametrization ($H_0=70.14 \pm 0.61$ km/s/Mpc). Thus, this decreases the tension in the absolute magnitude by 1.2$\sigma$, which is almost the same decrease in the Hubble tension level. In this sense, the absolute magnitude tension still holds although the Hubble tension is relaxed.

\begin{figure}
    \centering
    \includegraphics[width=0.6\textwidth]{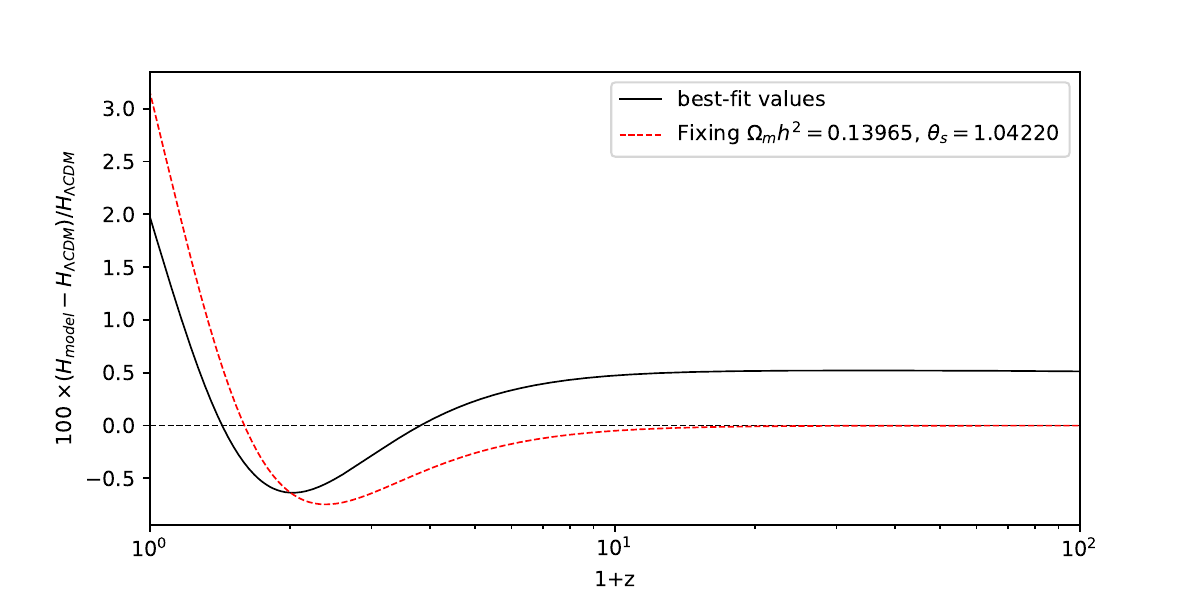}
    \caption{The percentage relative difference in the Hubble function between the Mukherjee model and $\Lambda$CDM  using \textit{all} dataset combination. We include two different scenarios: The first one (solid line) is by using the parameters' best-fit values as obtained in Tables \ref{tab:LCDM} and \ref{tab:MWN} for both models. The second one (dashed line) is by fixing the early universe to $\Lambda$CDM via fixing the acoustic scale $\theta_s = 1.04220$, and also the physical matter density $\Omega_mh^2 = 0.13965$. These are the best-fit values predicted by the $\Lambda$CDM.}
    \label{fig:Relative Difference in Hubble Function}
\end{figure}

We present the cosmic evolution via the Hubble function with respect to $\Lambda$CDM as seen in Figure \ref{fig:Relative Difference in Hubble Function}. The plots show the relative difference $\delta H/H_{\Lambda CDM}$ in percentage, where $\delta H \equiv H_{model} - H_{\Lambda CDM}$. The dashed line plot is obtained by fixing the early universe to $\Lambda$CDM via the acoustic scale $\theta_s$ and the matter density $\omega_m$, which fixes the cosmic expansion to $\Lambda$CDM at large $z$, only some deviation is obtained at small $z$. In this case, $\delta H$ is negative as expected in the phantom regime and becomes positive only at some $z<1$ to reach $H_0$ higher than $\Lambda$CDM by $\sim 3\%$ in their relative difference. However, this inevitably changes the distances at the late universe systematically and therefore will conflict with BAO angular distance \cite{El-Zant:2018bsc}. The solid line plot is obtained by allowing a tiny deviation $\sim 0.36\sigma$ in $\theta_s$ as obtained from the best-fit values of $\theta_s$ in Tables \ref{tab:LCDM} and \ref{tab:MWN} using \textit{All} combination dataset. In the latter case, we obtained $\delta H > 0$ at large $z$, it becomes negative at small $z$, whereas $\delta H$ is positive again at $z<1$ to address higher $H_0$ relative to $\Lambda$CDM by $\sim 2\%$. We note that the positive $\delta H$ values at $z \gtrsim 3$ partly compensate for negative $\delta H$ regions, which should alleviate the tension with late universe data. This will be tested in the next subsection.  

\begin{figure}
    \centering
    \includegraphics[width=0.85\textwidth]{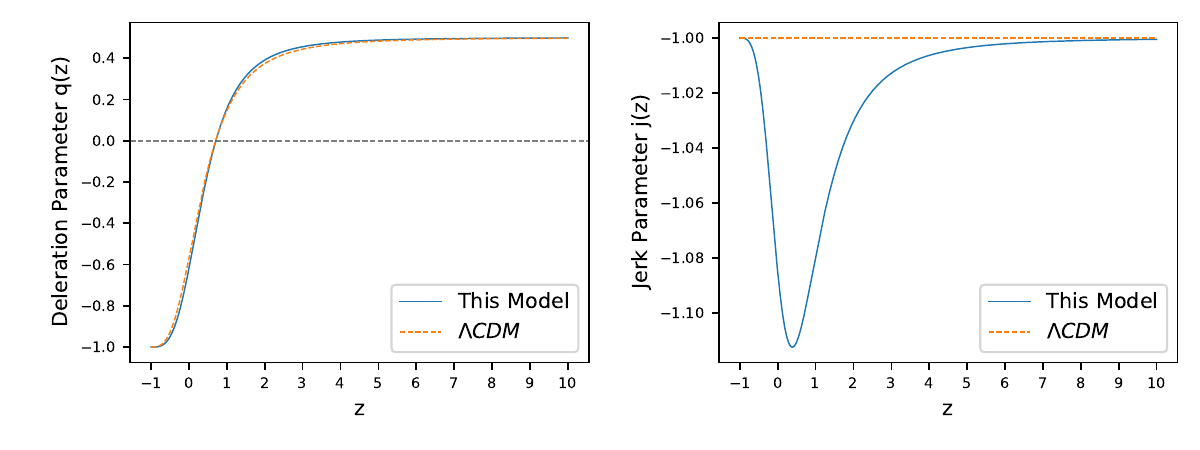}
    \caption{Evolution of the deceleration parameter, \textit{left}, and the jerk parameter, \textit{right}, with redshift z.}
    \label{fig:deceleration_and_jerk}
\end{figure}

Deceleration and jerk parameters evolution are shown in Figure \ref{fig:deceleration_and_jerk}. The parameters of the Mukherjee model agree with those of the $\Lambda$CDM model at the early universe epochs, while they deviate in the late universe. In the far future, $z\to-1$, the parameters are in agreement again in both models. At present, the model predicts a value for the jerk parameter $j_0 = -1.084\pm 0.029$ (68\% \textit{All}), which differs by $\approx 2.9\sigma$ from the constant value $j_0 = -1$ in $\Lambda$CDM. On the other hand, the deceleration parameter is found to be $q_0 = -0.618 \pm 0.023$ (68\% \textit{All}), which differs by $\approx 2.6 \sigma$ from $q_0 = -0.5556 \pm 0.0068$ (68\% \textit{All}) in $\Lambda$CDM. Notably, the jerk parameter in the present study differs by $\approx 2.1 \sigma$ from the previously obtained value $j_0 = -0.977 \pm 0.043$, while the deceleration parameter in the present study differs by $\approx 1.7\sigma$ from the previously obtained value $q_0 = -0.555 \pm 0.030$ based on background data \cite{Mukherjee_MWN}.
\subsection{Consistency of the Model with observational data}
\subsubsection{CMB and matter power spectra}
\begin{figure}
    \centering
    \includegraphics[width=0.45\textwidth]{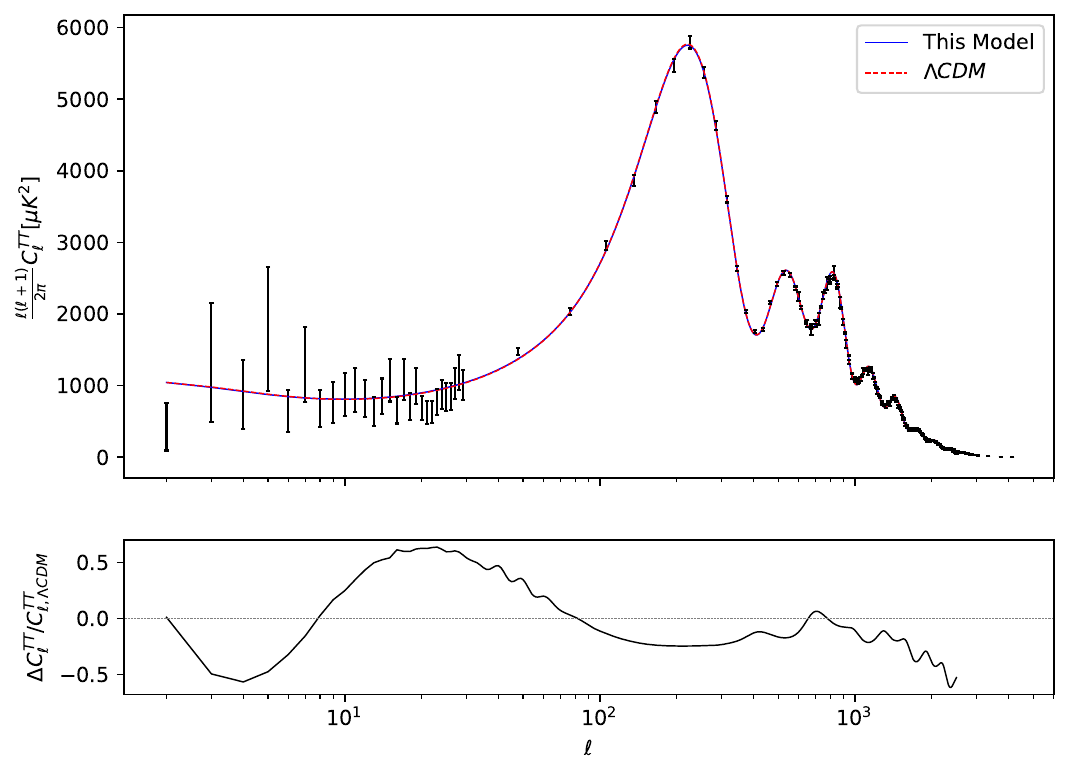}\hspace{10pt}
    \includegraphics[width=0.45\textwidth]{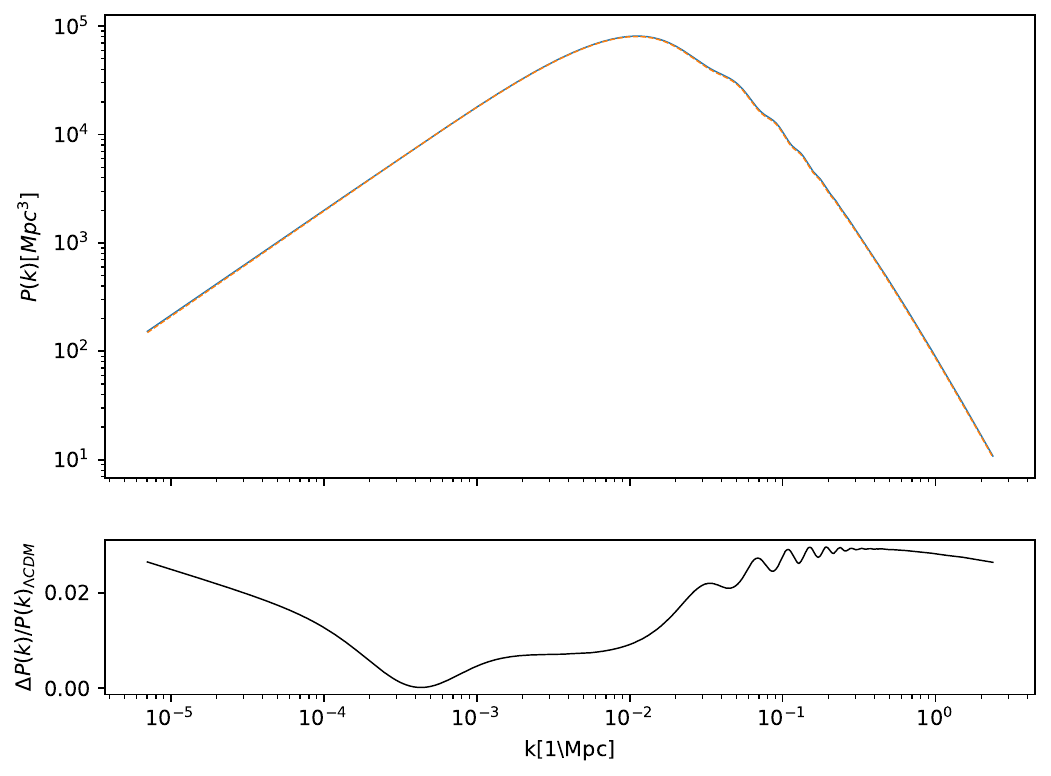}
    \caption{ \textit{Top left}: The CMB temperature power spectrum for both models by using the best-fit values of the parameters obtained from the combination of all datasets. \textit{Bottom left}: The percentage relative difference between the predictions of the model and $\Lambda$CDM. \textit{Top right}: The matter power spectrum $P(k)$ extrapolated to $z=0$ for both models using the best-fit values of the parameters obtained from the combination of all datasets. \textit{Bottom right}: The relative difference between the predictions of the model and $\Lambda$CDM.}
    \label{fig:powerspectra}
\end{figure}

From the previous discussion, we can see that the predictions of the six base parameters of the Mukherjee model agree at 68\% CL with $\Lambda$CDM for the various datasets mentioned. This agreement manifests in the CMB temperature power spectrum and the matter power spectrum as seen in Figure \ref{fig:powerspectra}.

Since dark energy is negligible in the early universe, compared to radiation and matter, we expect the evolution to be the same as predicted by $\Lambda$CDM and, hence, no deviation is expected between the two models in the CMB temperature anisotropy spectrum for large multipoles $(\ell \gtrsim  10^2)$. This explains the agreement between the model and $\Lambda$CDM, as shown in Figure \ref{fig:powerspectra} (left panel). The deviation between the two models at large multipoles we see in Figure \ref{fig:powerspectra} is due to the small fluctuations in the baryon and CDM energy density, in addition to the acoustic scale $\theta_s$. When using the same values for these parameters as in $\Lambda$CDM, these deviations vanish. For the small multipoles $(\ell \lesssim 10^2)$, the deviation is as small as 0.5\%, which is not measurable due to the large cosmic variance. This deviation is mainly due to the different behavior of dark energy in this model compared to $\Lambda$CDM, which manifests in the ISW effect.

In the right panel of Figure \ref{fig:powerspectra}, we plot the matter power spectrum at $z=0$ for Mukherjee and $\Lambda$CDM models. We can see the agreement between the two models, where the deviation from $\Lambda$CDM is within $\sim$0.025\% at all scales. This overestimated deviation in the matter power spectrum compared to $\Lambda$CDM explains the higher $\sigma_8$ value in the present model, see also the discussion on other local cosmological parameters in Appendix \ref{app1}. In conclusion, the model produces CMB and matter power spectra in agreement with Planck data.
\subsubsection{Late universe observations}
It proves convenient to confront the model with late universe data, i.e. SN, BAO, and growth rate of fluctuations. The main aim of this test is to examine any systematic deviations of the model prediction relative to observational data, which is associated with pure phantom dark energy as discussed earlier in this section.

\begin{figure}
    \centering
    \includegraphics[width=0.7\textwidth]{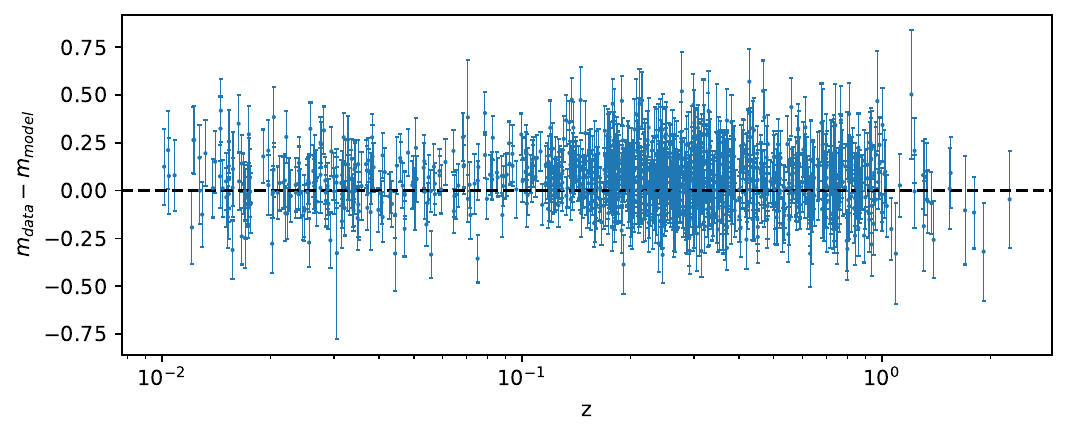}
    \caption{Deviation of Pantheon data from the model's prediction of the apparent magnitude $m$. The values of the parameters are the best-fit values obtained from the combination of all datasets.}
    \label{fig:luminosity_distance}
\end{figure}

We check the consistency between the model predictions and the Pantheon SNIa apparent magnitude $m_B$ data. The theoretical model contributes to the distance modulus $\mu$ where
\begin{equation}
    \mu(z) =  5\log\left(\frac{d_L}{10 pc}\right),
\end{equation}
with $d_L$ as the luminosity distance in Mpc. Therefore, the theoretical apparent magnitude $m_{th}$ is given by
\begin{equation}
    m_{th} = \mu + M,
\end{equation}
and $M$ is the SNIa absolute magnitude, which is a nuisance parameter, see Tables \ref{tab:LCDM} and \ref{tab:MWN}. In Figure \ref{fig:luminosity_distance}, we show the deviation between the measured Pantheon SNIa data points and the theoretically predicted values using the parameters' best-fit values from the combination of all datasets. A tiny shift of the data points of the Pantheon sample above the model prediction can be noticed in the figure. This is due to a tiny shift in the absolute magnitude when using the Pantheon sample. We note that the absolute magnitude is $M= -19.380 \pm 0.019$ when the Pantheon sample is used. This differs by $\sim 1.72\sigma$ from the present study when PantheonPlus is used.

\begin{figure}
    \centering
    \includegraphics[width=0.48\textwidth]{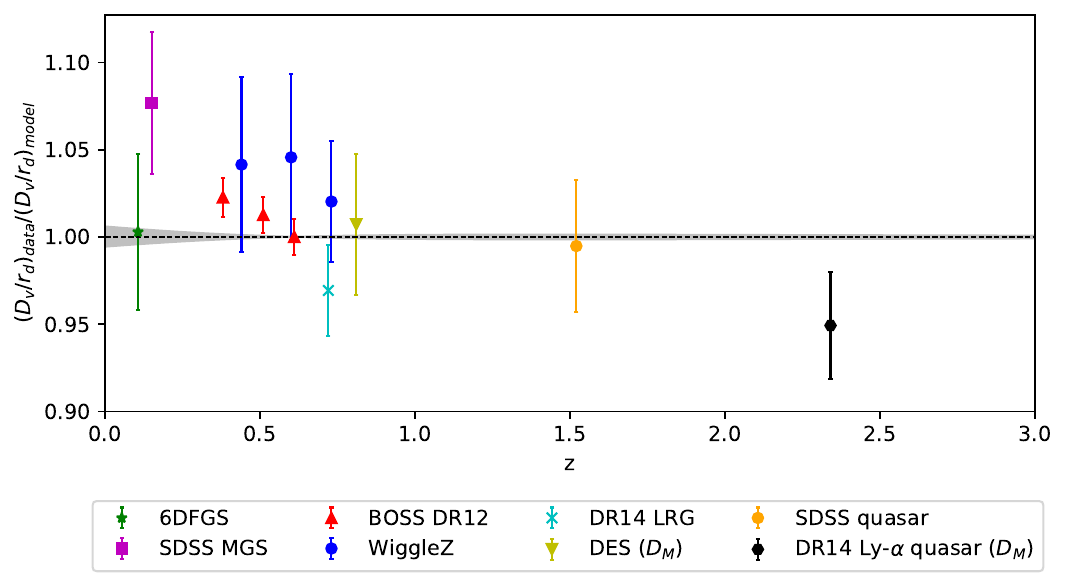}
    \includegraphics[width=0.48\textwidth]{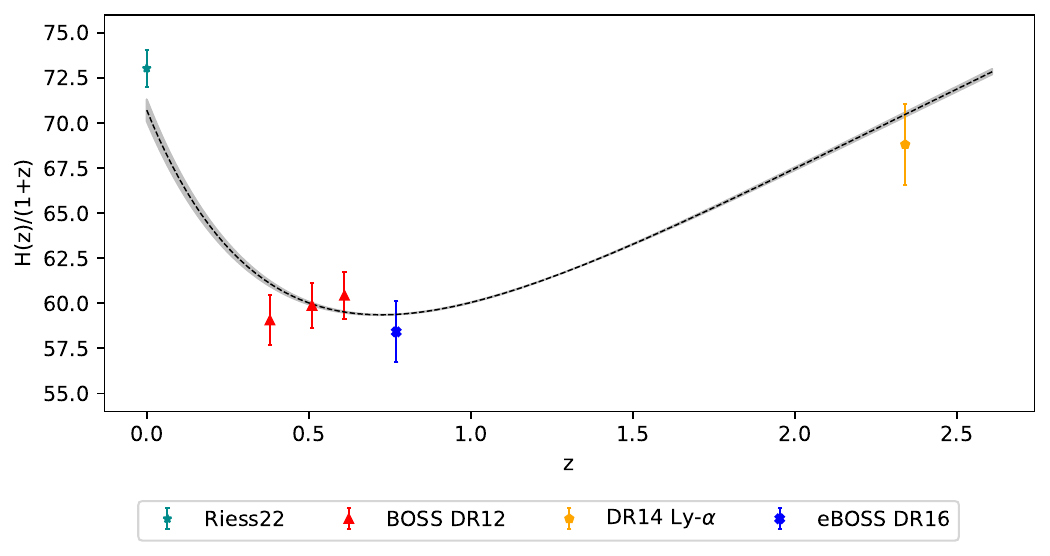}
    \caption{\textit{Left:} Acoustic scale distance measurements divided by the corresponding mean distance predicted by the model. The data used are as follows: 6dFGS \cite{BAO_6dF}, SDSS MGS \cite{MGS}, BOSS DR12 \cite{SDSS_DR12}, WiggleZ \cite{WiggleZ}, DR14 LRG \cite{DR14_LRG}, DES $(D_M)$ \cite{DES}, SDSS quasar \cite{DR14_quasar}, DR14 Ly-$\alpha$ quasar $(D_M)$ \cite{DR14_Ly_alpha}. \textit{Right:} The Hubble function evolution predicted by the Mukherjee model. The data used are as follows: Riess22 \cite{Riess2021}, BOSS DR12 \cite{SDSS_DR12}, DR14 Ly-$\alpha$ \cite{de_Sainte_Agathe_2019}, and eBOSS DR16 \cite{Wang_2020}. In both figures, we used the parameters' best-fit from the combination of all the datasets. The gray bands represent the $1\sigma$ confidence region.}
    \label{fig:BAO_data}
\end{figure}

We also checked the consistency between the model predictions and the BAO data from various measurements prior to the recent DESI BAO data. In Figure \ref{fig:BAO_data}, we plot the acoustic scale ratio $D_V(z)/r_d$ measured by various surveys divided by the predictions of the model at hand using the parameters' best-fit values using the combination of all datasets, where the gray band represents the $\pm1\sigma$ CL. The sound horizon at the drag epoch, characterized by the redshift $z_d$, is given by
\begin{equation}
    \label{r_d}
    r_d = \int^{\infty}_{z_d}\frac{c_s(z)}{H(z)} dz, \quad c_s(z) = \frac{1}{\sqrt{3(1+\mathcal{R})}}, \quad \mathcal{R} = \frac{3\Omega_b}{4\Omega_\gamma (1+z)}.
\end{equation}
The quantity $D_V(z)$ represents the isotropic evolution, which can be defined as
\begin{equation}
    \label{Dv}
    D_V(z) = [czD^2_M(z)/H(z)]^{1/3},
\end{equation}
where the comoving angular diameter distance $D_M$ is related to the physical angular diameter distance $D_A(z)$, by
\begin{equation}
    \label{ang.diam.dist}
    D_M(z) = (1+z) D_A(z), \quad D_A(z) = \frac{1}{1+z} \int^z_0 \frac{d\bar{z}}{H(\bar{z})}.
\end{equation}
Some systemic deviation of the data above the model predictions at low redshift $z<0.5$ can be noticed as shown by Figure \ref{fig:BAO_data}. As mentioned earlier in the previous subsection, allowing a small deviation in $\theta_s$ from $\Lambda$CDM leads to a change in the sign of $\delta H$ (see Figure \ref{fig:Relative Difference in Hubble Function}), which partly reduces the tension with late universe BAO distance measurements by fixing the early universe (fixing $\theta_s$ and $\omega_m$) to $\Lambda$CDM. Moreover, the dark energy EoS $w_{de}\sim -1.05$ is in the relevant redshift range, and therefore no large deviations in the BAO data are expected. We note that SDSS MGS and Ly-$\alpha$ measurements do not fit well the present model similar to the $\Lambda$CDM case.

\begin{figure}
    \centering
    \includegraphics[width=0.6\textwidth]{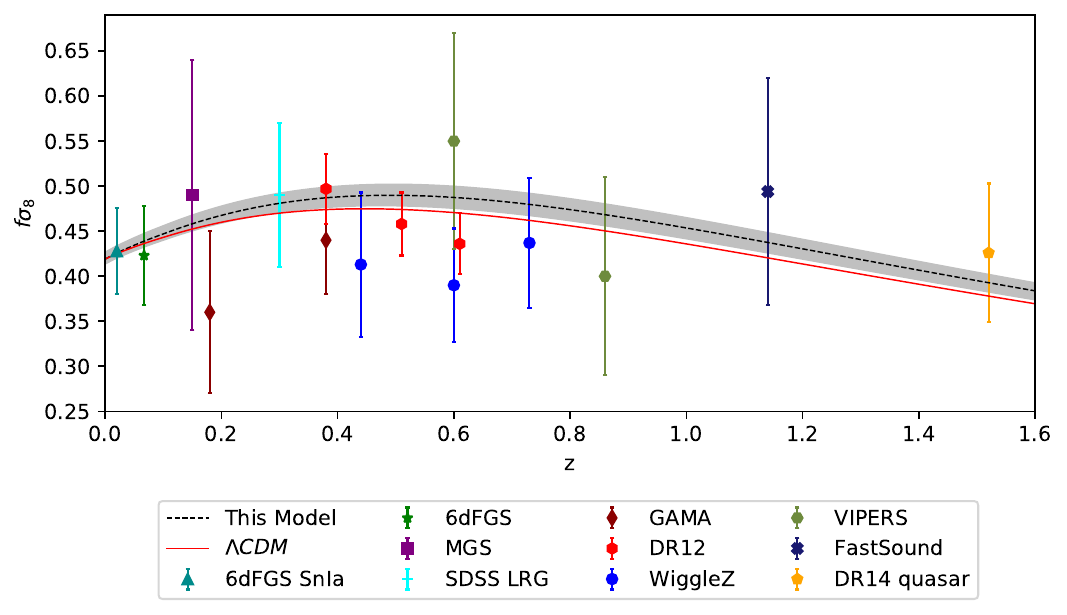}
    \caption{Growth rate of fluctuations predicted by $\Lambda$CDM and the model using the parameters' best-fit from the combination of all datasets. The data used are as follows: 6dFGS \cite{fsigma8_6dF}, 6dFGS SnIa \cite{fsigma8_6dF_SnIa}, MGS \cite{MGS}, SDSS LRG \cite{LRG_fsigma8}, GAMA\cite{GAMA_fsigma8}, DR12 \cite{SDSS_DR12}, WiggleZ \cite{WiggleZ_fsigma8}, VIPERS \cite{VIPERS_fsigma8}, FastSound \cite{FastSound_fsigma8}, DR14 quasar \cite{BOSS_DR14_quasar_fsigma8}. The gray band represents the $1\sigma$ confidence region for  the Mukherjee model.}
    \label{Fig:growthfn}
\end{figure}

Moreover, We check the consistency between the predictions of the model and the growth rate observational data. The evolution of the comoving matter density perturbation $\delta(a)$ at subhorizon scales can be written as \cite{Amendola_Tsujikawa_2010}
\begin{equation}
    \delta(a)''+\left(2+\frac{H'}{H}\right)\delta(a)'-\frac{3}{2} \Omega_m(a) \delta(a)=0
\end{equation}
where $'$ denotes $d/d\ln{a}$. The growth rate of fluctuations is often reported as the combination $f(a)\sigma_8(a)$, where $f(a)$ is the logarithmic growth rate of cosmological perturbations and $\sigma_8(a)$ is the normalization of the linear matter power spectrum at $a$. Those are defined as
\begin{equation}
    f(a) \equiv \frac{d\ln D(a)}{d\ln(a)},
\end{equation}
\begin{equation}
    \sigma^2_R(z) \equiv \langle\delta^2_{R}(x,z)\rangle = \int^\infty_0 \mathcal{P}(k,z) W^2(kR) d\ln k,
\end{equation}
where $D(a)\equiv \delta/\delta_0$ is the matter density contrast normalized to its current value $\delta_0$, $W(x)$ is the window function, $\mathcal{P}(k,z)$ is the dimensionless matter power spectrum, and $R = 8h^{-1} Mpc$. The evolution of $f\sigma_8(z)$ is plotted in Figure \ref{Fig:growthfn}. The growth rate as predicted by the present model is just above $\Lambda$CDM at all redshift range, but it is in agreement within $1\sigma$ CL with $\Lambda$CDM, and in general gives consistent predictions with the observational constraints at $1\sigma$ CL.

In conclusion, the above results reflect the agreement between the model predictions and observational late universe data. However, tiny systematic deviations from $\Lambda$CDM can be noticed with the matter power spectrum and late universe distance indicators.

\subsection{The Bayesian Evidence and Model Significance}
\label{Sec:Bayesian}
In order to test the model selection against $\Lambda$CDM for different datasets, we define $\Delta \chi^2_{min} \equiv \chi^2_\text{model}-\chi^2_{\Lambda CDM}$. In Table \ref{tab:Bayesian}, we calculate $\Delta \chi^2_{min}$ for different data combinations, where a negative value of $\Delta \chi^2_{min}$ indicates a preference of the model. In addition, we use Akaike Information Criteria (AIC), 
\begin{equation}
    \text{AIC}_{i}=\chi^2_{min,i}+2N_{i},
\end{equation}
to test the model against $\Lambda$CDM, where the number of free parameters of model $i$, $N_{i}$, are taken into account\footnote{The numbers of the free parameters are $N_{\Lambda CDM} = 6$ and $N_{model} = 7$ for $\Lambda$CDM and Mukherjee model, respectively.}. In Table \ref{tab:Bayesian}, the negative values of $\Delta\text{AIC}=\text{AIC}_{model}-\text{AIC}_{\Lambda CDM}$ indicate a preference of the model. We will follow Burnham $\&$ Anderson \cite{Burn2002} to quantify the evidence strength due to AIC. Accordingly, the model with $|\Delta \text{AIC}|$ between 0 and 2 is considered as substantial on the level of empirical support, while models with $|\Delta \text{AIC}|$ between 4 and 7 has considerably less support. For $|\Delta \text{AIC} > 10|$, the model may be omitted in future consideration.

We also used the publicly available code \texttt{MCEvidence} \cite{Heavens:2017afc,Heavens:2017hkr} to calculate the relative natural logarithmic Bayesian evidence of the present model compared to the reference $\Lambda$CDM model. We note that positive Bayes evidence, $\ln B_{ij}$, indicates a preference of the model over $\Lambda$CDM, where the strength of evidence for the model is quantified according to the revised Jeffreys' scale \cite{Kass:1995loi,Trotta:2008qt} as follows: $0\leq |\ln B_{ij}| < 1.0$ gives an inconclusive evidence, $1\leq |\ln B_{ij}| < 2.5$ gives a weak evidence, $2.5\leq |\ln B_{ij}| < 5$ gives a moderate evidence, $5\leq |\ln B_{ij}| \leq 10$ gives a strong evidence, and $|\ln B_{ij}| \geq 10$ gives a very strong evidence for the model. This applies in the opposite direction if $\ln B_{ij}$ is negative.
\begin{table}[ht]
\centering
    \caption{Model selection statistics, $\Delta \chi^2_{min} \equiv \chi^2_\text{model}-\chi^2_{\Lambda CDM}$, $\Delta\text{AIC}=\text{AIC}_{model}-\text{AIC}_{\Lambda CDM}$, where negative value indicates a preference for DDE model, and Bayes evidence $\ln B_{ij}$.
    The last two columns show the strength of evidence corresponding to Burnham and Anderson for $\Delta\text{AIC}$ and Jeffreys' scale for $\ln B_{ij}$.}
    \begin{adjustbox}{width=\textwidth}
    \begin{tabular}{l l c  c c c c c}
    \midrule
    \multirow{2}{*}
    {Model} &
    \multirow{2}{*}{Dataset} &
    \multirow{2}{*}{$\chi^2_{min}$} & \multirow{2}{*}{$\Delta \chi^2_{min}$} & \multirow{2}{*}{$\Delta AIC$} & \multirow{2}{*}{$\ln B_{ij}$ } & \multicolumn{2}{c} {Strength of evidence for the model}\\ & & & & & & $\Delta \text{AIC}$ & $\ln B_{ij}$ \\
    \midrule
     \multirow{4}{*}{$\Lambda CDM$} 
      & Planck \dotfill & $1012.97$ & --  & -- & -- & -- & --\\
      & Planck+DESI \dotfill& $1029.34$ & -- & -- & -- & -- & --\\
      & Planck+DESI+PantheonPlus \dotfill& $2441.62$ & -- & -- & -- & -- & --\\
      & Planck+DESI+PantheonPlus\&SH0ES \dotfill& $2352.10$ & -- & -- & -- & -- & -- \\
      \midrule
      \multirow{4}{*}{Mukherjee} 
      & Planck \dotfill& $1009.96$ & $-3.01$ & $-1.01$ & $+2.21$ &substantial & weak\\
      & Planck+DESI \dotfill& $1026.61$ & $-2.73$ & $-0.73$ & $+1.04$ & substantial & weak\\
      & Planck+DESI+PantheonPlus \dotfill& $2443.30$ & $+1.68$ & $+3.68$ & $-1.78$ & support ($\Lambda$CDM) & weak (in favor of $\Lambda$CDM)\\
      & Planck+DESI+PantheonPlus\&SH0ES \dotfill & $2346.16$ & $-5.94$ & $-3.94$ & $+1.80$ & support & weak\\
     \bottomrule  
    \end{tabular}
    \end{adjustbox}
    \label{tab:Bayesian}
\end{table}

In conclusion, all combinations, at 68\% CL, prefer phantom dark energy over cosmological constant, except for Planck + DESI + PantheonPlus which favors $\Lambda$CDM. We note that the latter dataset enforces the model towards $\Lambda$CDM, where the mean value of the model parameter $n=3.00\pm 0.11$. In this case, the model reproduces $\Lambda$CDM but with an extra free parameter, which explains the preference of $\Lambda$CDM. We would like to mention that the removal of data points at low-redshifts $z<0.01$ in PantheonPlus, to reduce the impact of peculiar velocities of the host galaxies, indicates a tension between local ($z<0.01$) and late universe ($z>0.01$) measurements. This can be seen clearly from Table \ref{tab:MWN}, which gives an absolute magnitude $M=-19.425\pm 0.018$. By adding PantheonPlus\&SH0ES instead, we obtain $M=-19.370\pm 0.015$. In effect, the latter gives higher $H_0$ than the former dataset, forcing the model towards a phantom regime. Indeed, the comparison between the first, second, and third rungs of SN groups in the PantheonPlus sample has been studied \cite{Huang:2024gfw}, where local-scale inhomogeneous new physics has been suggested to narrow down the $H_0$ tension \cite{Huang:2024erq}. In DESI DR2 using CPL, the results strengthen the evidence for the dynamical dark energy scenario when more data have been analyzed. Therefore, we expect DESI DR2 to offer a tighter constraint on the cosmological parameters in our case too, but it will not change the qualitative behavior.

Remarkably, the $w$CDM model introduces one free parameter similar to Mukherjee model but with fixed EoS. The BIC values of the $w$CDM model with respect to $\Lambda$CDM are as follows: For Planck dataset, $\ln B_{ij}=+2.79$, which provides a moderate evidence for the $w$CDM model. For Planck+DESI dataset, $\ln B_{ij}=-0.28$, which provides an inconclusive evidence. For Planck + DESI + PantheonPlus dataset, $\ln B_{ij}=-5.58$, which provides a strong evidence against the $w$CDM model. For Planck + DESI + PantheonPlus\&SH0ES dataset, $\ln B_{ij}=-0.23$, which provides an inconclusive evidence. In the Planck dataset (first case), the large cosmic variance at low-$\ell$ relevant to late DE era does not provide strict constraints on the model parameters. This has been already noticed by Planck collaboration \cite[see Sec. 7.4]{Planck18_params}. On the other hand, the second and the fourth cases provides no conclusive evidence for the $w$CDM model. The third case, gives a strong evidence against $w$CDM. This is in agreement with DESI results for $w$CDM where no evidence for a deviation from $w_{de}=-1$ has been found when a flat $w$CDM model is assumed \cite[see Sec. 5.1]{DESI}. Therefore, Mukherjee parametrization provides a novel model which cannot be covered by $w$CDM.

\section{Conclusions}
\label{Sec:Summary}

Several recent observational evidences may indicate tensions with the standard cosmological paradigm, i.e. $\Lambda$CDM model, where the dark energy is described by introducing a cosmological constant into Einstein's field equations. This is apart from the well-known longstanding theoretical disagreement on the cosmological constant value. Some of these tensions are at severe levels as in the famous $H_0$-tension which is at the level of $\sim 5\sigma$, while others are less severe like $\sigma_8$-tension which is at $\lesssim 3\sigma$. In addition, the latest DESI BAO observations show a preference for dynamical dark energy scenario. However, this conclusion strongly depends on the assumed priors on the CPL model, $w(a)=w_0+w_a(1-a)$, which introduces two free parameters $w_0$ and $w_a$.

Motivated by these evidences, we explored the Mukherjee effective equation of state which allows for both fixed and evolving dark energy behaviors \cite{Mukherjee_MWN}. This model introduces two parameters $n$ and $\alpha$ which have been tested using only background data. In the present work, we confronted the model with updated observational data, which allowed us to test the model on the background and linear perturbation levels. The model contains only one independent parameter $n$, since the other parameter $\alpha$ is completely determined by the matter density parameter $\Omega_{m,0}$ and $n$ via the relation $\alpha=\Omega_{m,0}^{n/3}\left(1-\Omega_{m,0}^{n/3}\right)^{-1}$. Interestingly, the model can perform distinguishable cosmic evolution scenarios according to some critical value of the model parameters $n$:
\begin{itemize}
    \item For $n=3$, the model produces $\Lambda$CDM scenario.
    \item For $n<3$, the model produces dynamical dark energy in \textit{quintessence} regime.
    \item For $n>3$, the model produces dynamical dark energy in \textit{phantom} regime.
\end{itemize}
Notably, the Mukherjee ``one-parameter" model reduces the free parameters by one in comparison to CPL parametrization which has been considered as the baseline in the DESI study. In this sense, a preference for the DDE scenario will be more robust than CPL which contains two parameters. We tested the model using Planck's CMB temperature, polarization, and lensing data, along with DESI, PantheonPlus SN, and PantheonPlus\&SH0ES data. This is achieved using \texttt{CLASS} \cite{CLASS}, \texttt{MontePython} \cite{MontePython}, and \texttt{GetDist} \cite{getDist}, ensuring that a Gelman-Rubing convergence criterion of $R-1 < 0.01$ is satisfied for all parameters \cite{Rubin_Gelman}. Combining all data enabled us to obtain $n = 3.30 \pm 0.11$ at 68\% CL, which favors a phantom dark energy behavior. We determined the current equation of state for dark energy $w_{de,0} = -1.047 \pm 0.016$.

We discussed the $n$-$H_0$ degeneracy. We showed that Planck data alone cannot set a strict constraint on $n$, where relatively higher $n$ values, greater than 3, are associated with more phantom DE regime and consequently increase $H_0$ values. This degeneracy can be removed by adding late-time observational data. \textit{Planck} dataset alone gives $n=4.16^{+0.81}_{-0.27}$ as well as $H_0=74.3^{+4.3}_{-1.6}$  km/s/Mpc, while the ($n, H_0$) parameter space is getting more confined when late-time data is included. We obtained $n=3.51^{+0.22}_{-0.28}$ and Hubble constant $H_0 = 70.9\pm 1.4$ km/s/Mpc, when including \textit{DESI} data. We note that this value decreases the tension with $H_0$ local measurements ($H_0=73.04\pm 1.04$ km/s/Mpc \cite{Riess2021}) to $1.2\sigma$, while $\Lambda$CDM is at $4.4\sigma$ tension and the CPL parametrization is at $3.7\sigma$ level \cite{DESI} using the same data. The combination \textit{Planck} + \textit{DESI} + \textit{PantheonPlus} gives more strict values $n=3.00\pm 0.11$ and $H_0 = 67.92 \pm 0.71$ km/s/Mpc, but reproduces $\Lambda$CDM scenario rather phantom DE. However, the inclusion of \textit{pantheonPlus\&SH0ES} puts the stringent constraint on the parameters where $n=3.30 \pm 0.11$ and $H_0 = 70.14 \pm 0.61$ km/s/Mpc. It is to be noted that \textit{Planck} + \textit{DESI} + \textit{PantheonPlus} combination is in tension with other datasets. We explained this tension due to the removal of the local universe ($z<0.01$) points in the PantheonPlus sample, which significantly affects the value of the absolute magnitude and reduces $H_0$ to the $\Lambda$CDM predictions. It is to be noted that all parameters have been reproduced using \texttt{Cobaya} sampling code with CMB lensing likelihoods from Planck+ACT, DESI, PantheonPlus, and PantheonPlus\&SH0ES. Only negligible shifts in the parameters are noticed, whereas all conclusions presented in this study remain the same.

We obtained slight changes from $\Lambda$CDM in the acoustic scale $\Delta \theta_s\sim 0.36\sigma$ and the matter density $\omega_m$, which derives $\delta H(z)$ to interpolate between positive and negative values along cosmic evolution. This helps to smooth the tensions with late time distance measurements from Pantheon SN, BAO prior to DESI, and the growth rate of cosmic structure. However, small systematic deviations from $\Lambda$CDM at late universe can be noticed, but consistent with the data. The BAO distance measurements of Ly-$\alpha$ forest with SDSS quasars and SDSS MGS disagree with the model similar to the $\Lambda$CDM model. We calculated the statistical parameters $\chi^2_{min}$, AIC, and the Bayesian evidence using \texttt{MCEvidence} public code, where all show a preference of the Mukherjee model for different combinations of datasets except the case when PantheonPlus is used. This can be sourced by the tension between local and late universe data.

In conclusion, the present model can perform different dark energy scenarios, where one scenario, or another, would be favored according to the observational constraints. Both AIC and BIC evidence analyses show a preference for DDE in phantom regime over cosmological constant DE when Planck, Planck + DESI, and Planck + DESI + PantheonPlus\&SH0ES datasets are used. 

\subsubsection*{Acknowledgment}
We would like to thank M. Hashim for continuous assistance during the preparation of this paper. We also would like to thank Eleonora di Valentino for valuable feedback and suggestions, which greatly improved the quality of this research. The work of E. I. Lashin is partially supported by Science, Technology \& Innovation Authority (STDF) under grant number 48173. 

\appendix
\section{Local universe cosmological parameters}
\label{app1}
\begin{figure}
    \centering
    \includegraphics[width=0.7\textwidth]{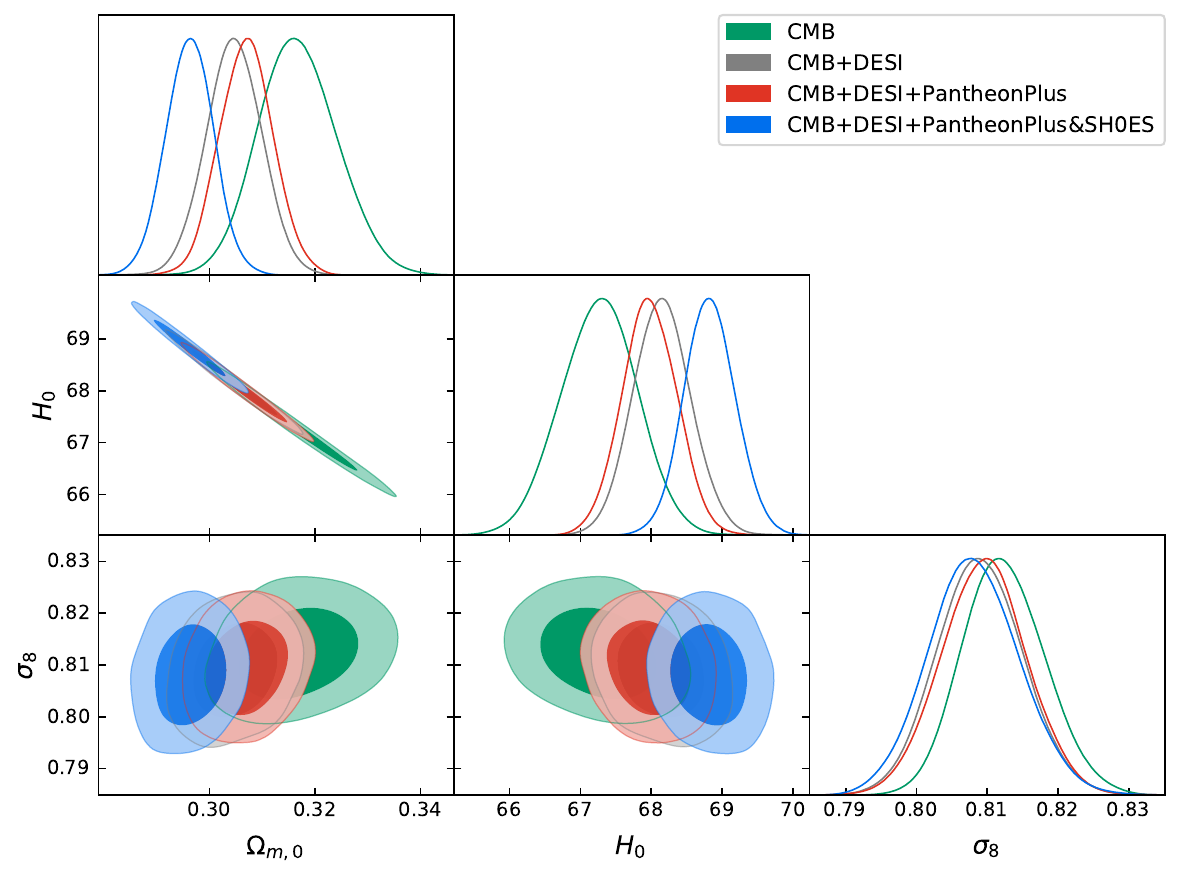}
    \caption{68\% and 95\% confidence contour plots of local cosmological parameters $H_0$, $\Omega_{m,0}$ and $\sigma_8$ in $\Lambda$CDM model for different datasets.}
    \label{Fig:LCDM_local}
\end{figure}

\begin{figure}
    \centering
    \includegraphics[width=0.9\textwidth]{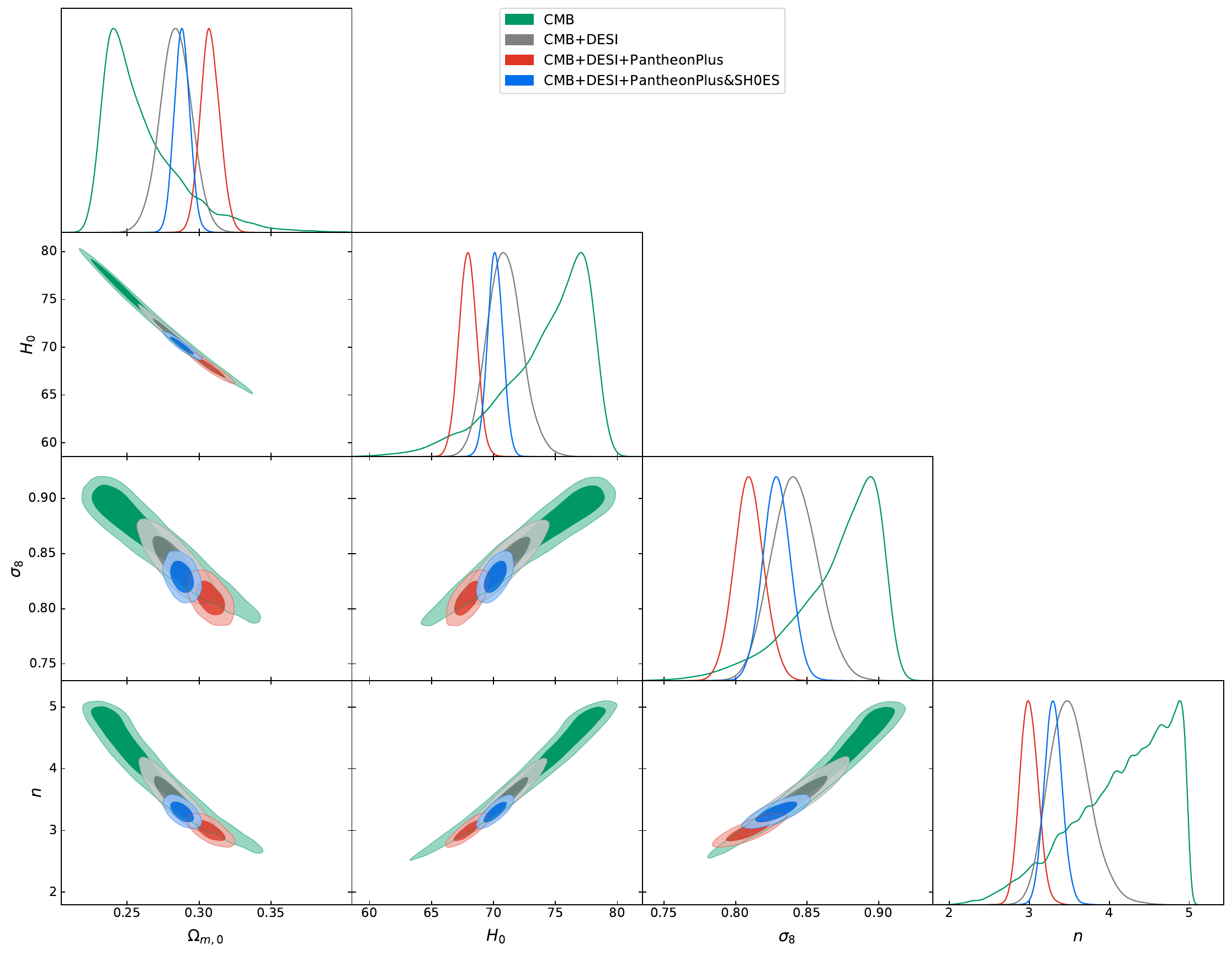}
    \caption{68\% and 95\% confidence contour plots of local cosmological parameters $H_0$, $\Omega_{m,0}$, $\sigma_8$ and Mukherjee EoS parameter $n$ for different datasets.}
    \label{Fig:MKH_local}
\end{figure}

In this section we discuss the local universe via the cosmological parameters $H_0$, $\Omega_{m,0}$ and $\sigma_8$. In Figure \ref{Fig:LCDM_local}, we show the 68\% and 95\% CL of the three parameters within $\Lambda$CDM for different data combinations. Similarly, in Figure \ref{Fig:MKH_local}, we show the same cosmological parameters in addition to the extra parameter $n$ associated with Mukherjee parametrization. Clearly, all datasets prefer phantom DE at 68\% as Mukherjee parameter $n>3$, only Planck+DESI+PantheonPlus data prefers the cosmological constant DE as $n=3$. This disagreement is due to removal of local SN data $z<0.01$ from the PantheonPlus sample. On the other side, Planck data alone weakly constrains the local parameters and the DE nature. This is due to large cosmic variance at low-$\ell$ relevant to late DE epoch. Adding late universe data, however, provides better constraints on the local universe cosmological parameters. 

We note that the redshift range at which the expansion rate $H(z)$ is lower in Mukherjee model than in $\Lambda$CDM (see Figure \ref{fig:Relative Difference in Hubble Function}), the growth rate $f(z)\propto \Omega_{m}(z)\propto (1+z)^3/H^2(z)$, is enhanced. However, at smaller redshift, when the Hubble rate crosses the corresponding $\Lambda$CDM rate to higher values including $H_0$ at $z=0$, then the growth rate should be suppressed including $\Omega_{m,0}$. This can be seen directly from the degeneracy relation $\Omega_{m,0}h^2$ which is fixed by the CMB power spectrum from Planck. In effect, the amplitude of the power spectrum $\sigma_8$ increases. This can be explained by the small overestimated deviation in the matter power spectrum $P(k)$ at $z=0$ of Mukherjee DE model in comparison to $\Lambda$CDM as seen in Figure \ref{fig:powerspectra} (Bottom right).
\begin{figure}
    \centering
    \includegraphics[width=0.65\textwidth]{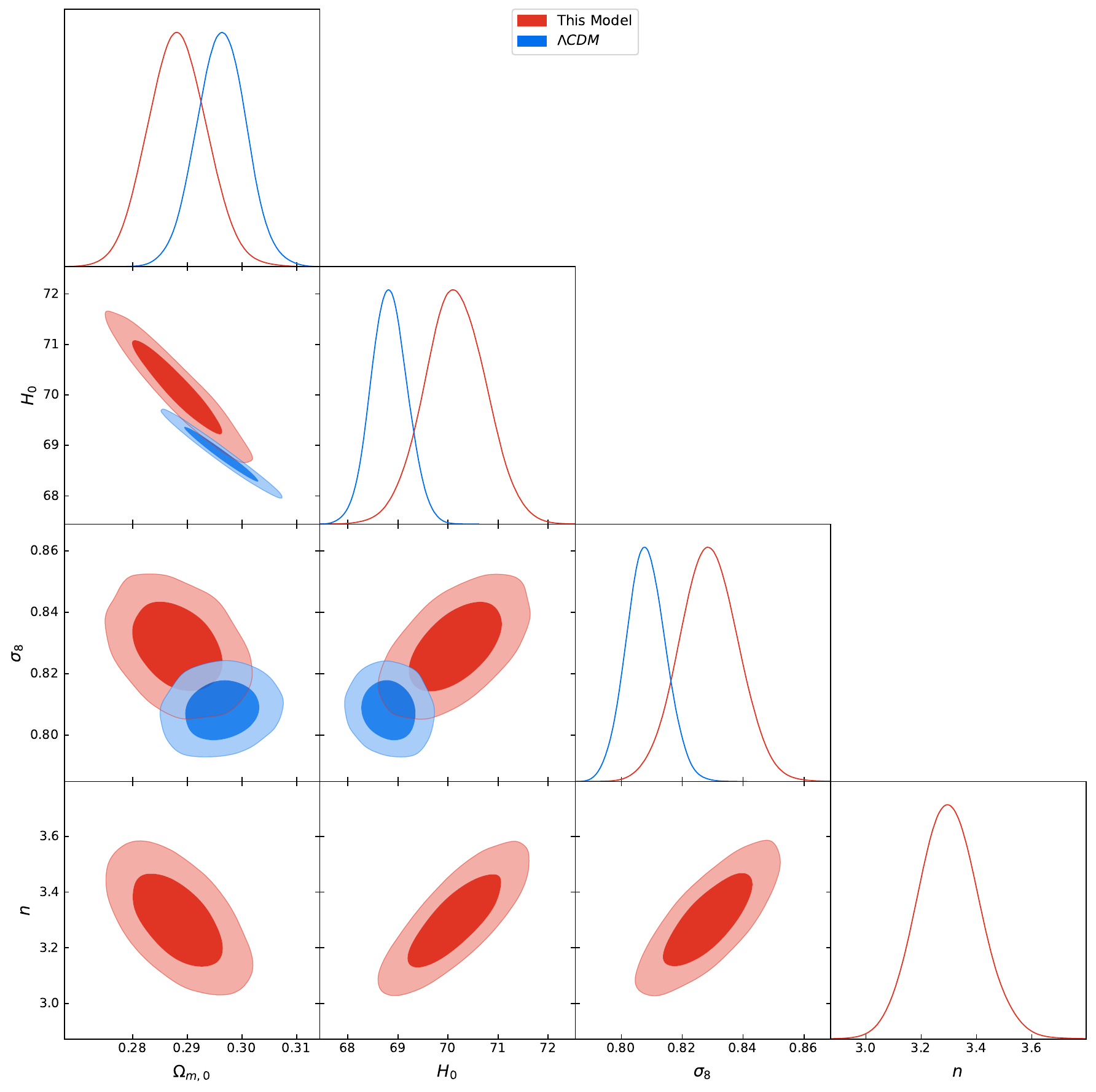}
    \caption{Comparison between 68\% and 95\% confidence contour plots of the $H_0$, $\Omega_{m,0}$, $\sigma_8$ in $\Lambda$CDM and Mukherjee model by including the EoS parameter $n$ using all dataset combination.}
    \label{Fig:derived_MKH}
\end{figure}

One may determine the performance of Mukherjee DE model in comparison with $\Lambda$CDM standard scenario via the local universe cosmological parameters. In Figure \ref{Fig:derived_MKH}, for the all dataset combination, we visualize the confidence contour plots of the local cosmological parameters $H_0$, $\Omega_{m,0}$ and $\sigma_8$ in $\Lambda$CDM ($n=3$) and in Mukherjee DE model by including the EoS extra parameter $n$. The figure shows clearly how the local universe parameters deviate from the standard $\Lambda$CDM scenario according to Mukherjee parametrization. The EoS parameter $n>3$ at $\gtrsim 2\sigma$ level which provides a phantom regime. Accordingly, as explained above, the local universe parameters $H_0$ and $\sigma_8$ have larger values in comparison to $\Lambda$CDM, while $\Omega_{m,0}$ is smaller.

\bibliographystyle{unsrtnat}
\bibliography{DDE_DESI}
\end{document}